\documentclass[12pt, a4paper]{article}
\usepackage{amsmath, amsthm, amsfonts}
\usepackage{geometry}
\usepackage[toc,page]{appendix}

\usepackage{tikz}
\usepackage[T1]{fontenc}
\usepackage{subfig}

\tikzstyle{level 1}=[level distance=3cm, sibling distance=1.0cm]
\tikzstyle{level 2}=[level distance=3cm, sibling distance=0.7cm]
\tikzstyle{level 3}=[level distance=3cm, sibling distance=0.4cm]

\usetikzlibrary{arrows}
\usetikzlibrary{trees}

\tikzset{
  treenode/.style = {align=center, inner sep=0pt, text centered,
    font=\sffamily},
  arn_n/.style = {treenode, circle, white, font=\sffamily\bfseries, draw=black,
    fill=black, text width=0.7em},
  arn_r/.style = {treenode, circle, red, draw=red, 
    text width=0.7em, very thick},
  arn_x/.style = {treenode, rectangle, draw=black,
    minimum width=0.95em, minimum height=0.95em}
}

\newtheorem{thm}{Theorem}[section]

\newtheorem{prop}[thm]{Proposition}
\theoremstyle{definition}

\theoremstyle{remark}

\numberwithin{equation}{subsection}
\numberwithin{figure}{subsection}

\title{On Integrated Chance Constraints in ALM for Pension Funds}
\author{Youssouf A. F. Toukourou\footnote{Dept. of Actuarial Science, Faculty of Business and Economics, University of Lausanne UNIL-Dorigny, CH-1015 Lausanne, Switzerland.} \footnote{\underline{email:} youssouf.toukourou@unil.ch} $\;$and Fran{\c c}ois Dufresne$^{*}$\footnote{\underline{email:} francois.dufresne@unil.ch} \\
}

\begin{document}
\maketitle

\abstract{We discuss the role of \textit{integrated chance constraints} (ICC) as quantitative risk constraints in asset and liability management (ALM) for pension funds. We define two types of ICC: the \textit{one period} integrated chance constraint (OICC) and the \textit{multiperiod} integrated chance constraint (MICC). As their names suggest, the OICC covers only one period whereas several periods are taken into account with the MICC. A multistage stochastic linear programming model is therefore developed for this purpose and a special mention is paid to the modeling of the MICC.
\\Based on a numerical example, we firstly analyse the effects of the OICC and the MICC on the optimal decisions (asset allocation and contribution rate) of a pension fund. By definition, the MICC is more restrictive and safer compared to the OICC. Secondly, we quantify this MICC safety increase. The results show that although the optimal decisions from the OICC and the MICC differ, the total costs are very close, showing that the MICC is definitely a better approach since it is more prudent.}
\\
\\
\\
\\
\\
\\
\\
\\
\\
\\
\\
\\
\\
\\
\\
\\
\pagebreak

\section{Introduction}
A pension fund is any plan, fund or scheme, established by a company, governmental institution or labour union, which provides retirement incomes. The actuarial present value of current and future payments constitutes the total liability of the fund. The pension fund receives contributions from its active members and/or the employer. This money (considered as the total wealth or total asset) is invested in a wide range of assets. The asset allocation is made in such a way that it guarantees, to a certain extent, the payments of future obligations. That is not so trivial: assets yield random returns and future benefits are not known with certainty. An asset liability management study (ALM) provides a rich theoretical background to address that issue. Its goal is to determine the adequate asset allocations and contribution rates in order to guarantee the payment of current and future pensions.
\\The use of ALM methods has a long tradition in pension funds. At the beginning, it has started with \textit{deterministic methods}. According to those methods, the future cash flows are estimated and assumed to be certain; the wealth is mainly allocated to bonds considered as risk free. Bonds are chosen in such a way that their related incomes correspond to yearly pension payments. These models are essentially based on immunization and cash-flow matching; see, for example, Koopmans \cite{koopmans1942risk} and Redington \cite{redington1952review}. Deterministic methods have proven to be inefficient since uncertainty turns out to be increasingly difficult to handle. The problem of uncertainty is somehow taken into account by \textit{stochastic methods}, especially by the way of \textit{surplus optimization theory}. This approach is often based on the efficient portfolio theory of Markowitz \cite{markowitz1952harry}. The literature has flourished in that field and we can cite Sharp and Tint \cite{sharpe1990liabilities} and Leibowitz \cite{leibowitz1992asset} among others. However, the pension fund problem is a long term problem with a horizon span of approximatively 30 years. Hence, its model should be dynamic. Furthermore, regulations often impose many types of constraints. Those matters are hardly taken into account by surplus optimization methods. In practice, \textit{simulation methods} are commonly used due to their ability to incorporate the above issues. Initially, they consisted on defining a set of feasible allocations and contribution rates, and chosing the best one in some sense. The choice is based on the simulation of the future paths. Due to the technical innovations, these methods have significantly evolved with the work of Wilkie \cite{wilkie1995more} and Ahlgrim and al. \cite{ahlgrim2005modeling} concerning economic scenario generation. M{\o}ller and Steffensen \cite{moller2007market} provide different tools for valuing the pension fund liabilities. Recent years have also seen the emergence of methods known as \textit{stochastic programming}.
\\Often based on simulations due to its complexity, stochastic programming gives a flexible and powerful tool for ALMs. Its importance lies in its ability to bring together many kinds of features in a common framework. Moreover, assets and liabilities are all influenced by many sources of risk and the risk aversion is accommodated; the framework has a long time horizon split into subperiods (\textit{multistage}); the portfolio can be rebalanced dynamically at the beginning of each subperiod; all these are incorporated in a single and consistent structure while satisfying operational or regulatory restrictions and policy requirements. Multistage stochastic programs (MSP) models have been applied to ALM for pension funds by Carino and al. \cite{carino1994russell}, Consigli and Dempster \cite{consigli1998dynamic}, Kusy and Ziemba \cite{ziemba1998worldwide} and Kouwenberg and Zenios \cite{kouwenberg2006stochastic}.
\\The ALM model in this paper is a MSP, for which, we minimize the total funding cost under risk, legal, budget, regulatory and operating constraints. The total funding cost is composed of regular and remedial contributions. Regular contribution constitutes a certain proportion (contribution rate) of the total salary whereas remedial contribution is an additional financial support provided by the employer (or a sponsor) whenever the solvency target is in question. More specifically, we  focus on the risk constraints, which are of \textit{integrated chance constraints} (ICC) type in this work. As an alternative to chance constraints (CC), ICC is computationaly of great interest; in particular when a quantitative risk measure is preferable. We define the \textit{funding ratio} as the ratio of total asset over total liability. Our goal is to meet a certain funding ratio, called here \textit{target funding ratio}, at the end of each subperiod. For a predefined target funding ratio, the ICC put an upper bound on the expected shortfall, i.e. the expected amount by which the goal is not attained. Haneveld and al. \cite{haneveld2010alm} and Drijver \cite{drijver2005asset} pionnered the application of ICC in ALM for pension fund. However, the risk parameter considered in their models is neither scale free, nor time dependent. Our model is close to Haneveld and al. \cite{haneveld2010alm} with the particularity that the risk parameter is a linear function of total liability. Then, it becomes unvariant with respect to the size of the fund as well as time dependent.
\\We define two types of ICC: the \textit{one period} integrated chance constraint (OICC) and the \textit{multiperiod} integrated chance constraint (MICC). As their names suggest, the OICC covers only one period whereas several periods are taken into account with the MICC. A multistage stochastic \textit{linear} program is therefore developed for this purpose and a special mention is paid to the modeling of the MICC.
\\The rest of the paper is organized as follows. In section \ref{setting}, the theoretical background, the dynamics and the ALM optimization problem are extensively detailed. Section \ref{Framework} defines the risk contraints and shows how CC leads to ICC. Moreover, OICC and MICC are introduced and their stochastic linear program reformulations are derived as well. In section \ref{Num_ill}, a numerical example is examined from the perspective of a defined benefit fund that invests in stocks, real estate, bonds, deposits and cash. All numerical results are implemented using the solver CPLEX in the mathematical programming language AMPL. We first analyse the effect of the risk parameter on the optimal decisions. This section finishes by a brief comparison of the two ICC. Section \ref{conclusion} concludes the paper.

\section{Settings}\label{setting}
This chapter introduces the dynamics of the ALM model with its specific features.

\subsection{Multistage recourse models}\label{multistage recourse models}
\begin{figure}[t!]
\centering
\begin{tikzpicture}[grow=right,->,>=stealth'] 
\node [arn_n] {}
    child{ node [arn_n] {} 
            child{ node [arn_n] {}
							child{ node [arn_n] {{\tiny 40}}}
							child{ node [arn_n] {{\tiny 39}}}
            } 
            child{ node [arn_n] {}}
            child{ node [arn_n] {}}
            child{ node [arn_n] {} 
            	        child{ node [arn_n] {{\tiny 34}} edge from parent node[above left]
                         {}} 
							child{ node [arn_n] {{\tiny 33}}}
            }                                       
    }
    child{ node [arn_n] {}}
    child{ node [arn_n] {}}
    child{ node [arn_n] {}}
    child{ node [arn_r] {}
            child{ node [arn_n] {}
                    child{ node [arn_n] {{\tiny 8}}}
				    child{ node [arn_n] {{\tiny 7}}}} 
            child{ node [arn_n] {}}
            child{ node [arn_n] {}}
            child{ node [arn_n] {}
							child{ node [arn_n] {{\tiny 2}}}
							child{ node [arn_n] {{\tiny 1}}}
            }
		}
; 
\end{tikzpicture}
${\tiny t=0}\;\;\;\;\;\;\; \text{event} \;\;\;\;\;\;\;{\tiny t=1}\;\;\;\;\; \text{event} \;\;\;\;\;{\tiny t=2}\;\;\;\;\; \text{event} \;\;\;\;\;{\tiny t=3}$
\caption{\label{arbre_scenario}A scenario tree with $40$ scenarios and $66$ nodes.}
\end{figure}
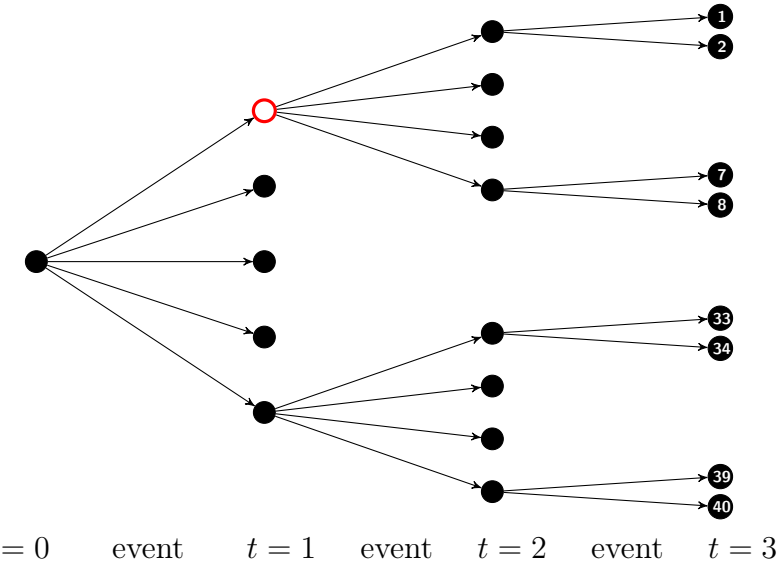

In this section, we describe the classical architecture of the multiperiod decision framework. The model's setup presented here ressembles mostly to Haneveld and al. \cite{haneveld2010alm}.
\\Since we aim for strategic decisions, we model the ALM process over a number of years and one set of decisions is taken each year. We discretize time accordingly so that the model has a (finite) number of one-year time periods. Consequently, we assume that the ALM model has a horizon of $T$ years from now, split in $T$ subperiods of one year each. The resulting years are denoted by an index $t$, where time $t=0$ is the current time. By year $t\;\left(t= 1,\cdots,T\right) $, we mean the span of time $\left[ t-1,t\right) $. We define
\begin{equation}
\mathcal{T}_t:=\left\lbrace t,t+1,\cdots,T\right\rbrace .
\nonumber
\end{equation}
We assume that uncertain parameters (e.g. asset returns) can be modeled as random variables with known distributions. At each time $t \in \mathcal{T}_0$, the pension fund is allowed to make decisions (corresponding to yearly corrections), based on the actual knowledge of parameters. During each one-year period, a realization of the corresponding random parameters becomes known (e.g. assets return during that year). That is, the concept underlying our model is the following sequence of decisions and observations:
{\center decide\;\;\;\;\;\;\;\;\;observe\;\;\;\;\;\;\;\;\;decide\;\;\;\;\;\;\;\;\;\;\;\;\;\;\;\;\;\;\;\;\;\;\;\;\;\;\;\;\;\;observe\;\;\;\;\;\;\;\;\;\;\;\;decide\;\;\;\;\;\;\;\;\;\;\;\;observe
\\ $ \;\;\;\;\; X_0 \;\;\;\;\;\leadsto \;\;\;\;\; \omega_1 \;\;\;\;\;\leadsto \;\;\;\;\;X_1 \;\;\;\;\;\leadsto \;\;\;\;\;\cdots\;\;\;\;\;\leadsto\;\;\;\;\; \omega_{T-1} \;\;\;\;\;\leadsto \;\;\;\;\;X_{T-1} \;\;\;\;\;\leadsto \;\;\;\;\; \omega_T $\newline
\\where $X_t$ is the vector of decision variables at time $t\in\mathcal{T}_0$, and vector $\omega_t$, $t\in\mathcal{T}_1$ models all economic events which are the source of uncertainty and risk for the pension fund management, which, in our case, are asset returns as well as random contributions and liability streams.} Time $t$ is assumed to be the end of the financial year $t$. We assume that a financial year coincides with a calendar year. At time $t\in\mathcal{T}_0$, decisions $X_t$ are taken with full knowledge of the past $\left[ 0,t\right] $ but with only probabilistic informations about the future $\left( t,T\right] $.
\\Uncertainty in the model is expressed through a finite number $S$ of sample paths spanning from $t=0$ until $t=T$ called scenarios. That is, we assume the random variable follows a discrete distribution with $S$ possible outcomes. Each scenario represents a sequence of possible realizations of all uncertain parameters in the model. As explained above, $\omega_t$ is the stochastic vector process whose values are revealed in year $t$. Then, the set of all scenarios is the set of all realizations $\omega^s := \left( \omega_1^s,\cdots,\omega_T^s \right), s\in \mathcal{S}:=\left\lbrace 1,\cdots,S\right\rbrace  $ of $\omega := \left( \omega_1,\cdots,\omega_T \right)$. Scenario $s$ has a probability $p^s$, where $p^s>0$ and $\sum_{s=1}^S p^s =1$. It represents a description of possible future, starting just after $t=0$. If we assume that we can observe the "state of the world" at time $t, \left( 0<t<T\right) $, then there is a unique history of realizations of $\left( \omega_1,\cdots,\omega_{t-1} \right)$ leading to that state, but the future as seen from time $t$ may unfold in several ways. That is, there are several distinct scenarios which share a common history up to time $t$. A suitable representation of the set of scenarios is given by a scenario tree (see Figure $\ref{arbre_scenario}$). In respect to Figure $\ref{arbre_scenario}$, we define the node as the possible outcome of the stochastic vector $\omega_t$ at a given time $t \in \mathcal{T}_0 $. Each path of $\omega_t$ from $t=0$ to $t=3$ represents one scenario; each node of the scenario tree has multiple sucessors, in order to model the process of information being revealed progressively through time. By convention, the scenarios are numbered top-down by their end node. The arcs in the tree denote realizations in one time period. We assume here that, for a specific decision time $t\in\mathcal{T}_0$, the numbers of realizations in one time period descending from the current nodes are identical.
\\For example in Figure $\ref{arbre_scenario}$, we have a $3$-year horizon scenario tree with $40$ scenarios. Over the first period starting from time $0$ to time $1$, there are five possible realizations. From each of these realisations, we have four possible outcomes over the second year; each of them is a conditionnal realisation as it depends on the preceding node. Over the third period, each of the second period observations can lead to two possible outcomes. All this gives a branching structure of $1-5-4-2$ and leading to a total of $S=5\times4\times2=40$ possible scenarios.
\\A multistage recourse model is an optimization problem defined on such a scenario tree. Considering the remaining future represented by the subtree rooted at $\left( t,s\right) $, optimal decisions are taken for each node $\left( t,s\right) $ of the event tree, given the informations available at that point. Optimality is defined in terms of current costs plus expected future costs, which are computed with respect to the appropriate conditional distributions, Vlerk and al. \cite{van2003integrated}.\\
\\Ideally, one would like to make different decisions for every path at every $t\in \mathcal{T}_0$, but this would lead to undesirable anticipativity in the model. The simplest way to avoid this is to make one single decision at each time $t$ for all paths by adding explicit constraints. That is, for any two different scenarios $s_1$ and $s_2$ ($s_1,s_2 \in \mathcal{S}$ and $s_1 \neq s_2$) having the same history up to time $t\in \mathcal{T}_0$, we enforce $X_t^{s_1}=X_t^{s_2}$, where $X_t^s$ is the decision $X_t$ under scenario $s$. For example, at the empty circle of Figure $\ref{arbre_scenario}$, $ X_1^1 = X_1^2 = \cdots = X_1^8 $.

\subsection{Dynamics}\label{Dynamics}
\subsubsection{Assets} \label{assets}
In this paper, we are considering a buy and hold model applied to a DB plan in which one seeks to minimize the expected cost of funding. In this respect, dynamics for both assets and liabilities should clearly be specified.
\\At initial time $t=0$, the exact levels of wealth and liability are available to the decision maker who has to decide, each period, how to rearrange his portfolio in order to cover liabilities and, at the same time, to achieve high returns on the financial market. The higher the returns are, the lower the contribution rate could be. Let denote $A_t$ the total amount of wealth at time $ t\in\mathcal{T}_0 $. The total wealth is allocated into $d$ classes of assets and in cash. Let $k\in\mathcal{K}:=\left\lbrace 1,...,d\right\rbrace $ denote the asset class index. At each decision time $ t\in\mathcal{T}_0$, a specified amount of $H_{k,t}$ is allocated to asset $k$ and $C_t$ is the cash amount. We can write
\begin{equation}
A_t=\sum_{k=1}^{d}H_{k,t} + C_{t}.
\label{budget_constraint}
\nonumber
\end{equation}
Through buying and selling, the investor restructures his portfolio at each time $t$. Once the $t^{th}$ stage decision is made, the holdings $H_{k,t}$ can be calculated. The shares in the portfolio are then kept constant till the next decision time. The value of $H_{k,t}$ is affected by the returns on the market. Let define $\xi_{k,t}:=1+r_{k,t}$ where $r_{k,t}$ is the random rate of return on asset class $k$ over year $t$.
\\Over year $t$, the pension fund pays benefits to its non-active members and receives contributions from its active members or/and the employer (also called the sponsor). Benefits regroup pensions which are paid to retirees, disability and death annuities or lump sum, whereas contributions are composed of yearly payments from all the active members and/or sponsor to the plan. When it appears that the plan is unfunded according to its solvency target, the sponsor may finance the deficit. As in Vlerk and al. \cite{van2003integrated}, we name this funding here as remedial contribution. We then assume that whenever the solvency target is not fulfilled, a remedial contribution in cash can be obtained from the sponsor. In practice, it does not really work that way. For example in Vlerk and al. \cite{van2003integrated}, the remedial is only provided after two consecutive periods of underfunding. We will see in the model description that the parameters are set such that the remedial contribution variable is non-zero only under some conditions. In general for DB plans, future benefits and liabilities depend on company policy regulation and can be estimated whereas yearly contribution is defined as a certain proportion of the yearly salary. Asset allocation and contribution rate are defined with respect to the level of future benefits and liabilities (e.g. Switzerland). Kim \cite{kim2008morneau} provides a rich source of informations concerning different types of pension plans and features. During year $t$, let $\text{Ben}_t$ and $W_t$ denote, respectively, the total amount of benefits paid and the level of salary. The variable $cr_t$ is the decided contribution rate for year $t+1$. For returns and cash-flow variables, index $t$ means that payments occur over year $t$ but cash-flows are accounted at the end of year. Accordingly, the total asset dynamic is modeled as
\begin{equation}
A_{t}=\sum_{k=1}^{d}H_{k,t-1}\xi_{k,t} + C_{t-1}\left( 1+r_f\right)  + cr_{t-1}W_t-\text{Ben}_t+ Z_t =\sum_{k=1}^{d}H_{k,t}+C_t,
\label{asset_dynamic}
\end{equation}
for $t\in \mathcal{T}_1$, where $r_f$ is the risk free interest rate and $Z_t$ is the remedial contribution at time $t$. Before receiving the remedial contribution at time $t$, the total wealth is defined as $A^*_t$ and one can write
\begin{equation}
A^*_{t}=\sum_{k=1}^{d}H_{k,t-1}\xi_{k,t} + C_{t-1}\left( 1+r_f\right)  + cr_{t-1}W_t-\text{Ben}_t =A_t - Z_t.
\label{asset_dynamic_00}
\end{equation}
Asset returns ${\left(\xi_{t}\right) }_{t=1}^T := \begin{pmatrix}
{\xi_{1,t} } &\cdots & {\xi_{k,t} } & \cdots & {\xi_{d,t} }
\end{pmatrix}_{t=1}^T$, pension payments $Ben_t$ and salary $W_t$ are modeled as stochastic processes on a filtered probability space $ \left(\Omega,\mathcal{F},{\left( \mathcal{F}\right) }_{t=1}^T,\mathrm {P}\right)   $. Obviously, at each decision time, $A_t$ is a random variable whose distribution depends, on a first hand, on $\xi_t$, $W_t$ and $\text{Ben}_t$, and on the second hand, on asset allocations before $t$. At a specific date $t$, the variable $A_t$ is known as it can be observed. The initial wealth is defined by $\bar{A_0}$  and known at initial time $t=0$. According to \eqref{asset_dynamic}, total wealth $A_{t-1}$ at time $t-1$  is allocated into the $d$ classes of assets and cash. Each asset $k$ generates a return $\xi_{k,t}$ over period $\left[ t-1,t\right]$. The initial wealth plus accumulated interest at the end of period will be augmented by the balance of external flows: contributions minus pension payments. This latter can be either positive or negative depending on the difference between contributions and benefits paid. A negative balance could be due to the fact that a company is no more hiring new employees. This may happen for various reasons: runoff, economic difficulties, etc. Obviously, the total contributions $cr_{t-1}W_t$  will decrease considerably as the total salaries decrease whereas $\text{Ben}_t$ will tend to increase as people leave the fund. When at time $t$ the total asset can not fulfill the pension fund solvency target, it may obtain a remedial contribution $Z_t$.
\\In order to be as close as possible to realities on the financial market, one has to consider costs of trading activities. Therefore, we include proportional transaction costs $\bar{c}^B:=\begin{pmatrix}
\bar{c}_1^B & \ldots & \bar{c}_k^B & \ldots & \bar{c}_d^B
\end{pmatrix}$ and $\bar{c}^S:=\begin{pmatrix}
\bar{c}_1^S & \ldots & \bar{c}_k^S & \ldots & \bar{c}_d^S
\end{pmatrix}$ for purchases and sales, respectively. Inclusion of transaction costs will lead to some changes in asset dynamics. Thus, \eqref{asset_dynamic} is then replaced by
\begin{equation}
A_{T}=\sum_{k=1}^{d}H_{T-1,k}\xi_{T,k} + C_{T-1}\left( 1+r_f\right)  + cr_{T-1}W_T-\text{Ben}_T=A_T^{*}
\label{asset_dynamic_1_}
\end{equation}
over period $\left[ T-1,T\right]$ and when $ t\in\mathcal{T}_1 \setminus \left\lbrace T \right\rbrace  $,
\begin{equation}
\begin{split}
A_{t} & =\sum_{k=1}^{d}H_{k,t-1}\xi_{k,t} + C_{t-1}\left( 1+r_f\right)  + cr_{t-1}W_t - \text{Ben}_t + Z_t  - \sum_{k=1}^{d} \left( \bar{c}_k^BB_{k,t}+\bar{c}_k^SS_{k,t}\right)
\\ & =\xi_{t}H_{t-1}+C_{t-1}\left( 1+r_f\right) + cr_{t-1}W_t - \text{Ben}_t + Z_t  -\left( \bar{c}^B B_{t}+\bar{c}^S S_{t}\right)
\\ & = A_t^{*}+Z_t -\left( \bar{c}^B B_{t}+\bar{c}^S S_{t}\right)
\\ & =\mathbf{e} \cdot H_t + C_t
\label{asset_dynamic_1}
\end{split}
\end{equation}
where $\mathbf{e}:= \begin{pmatrix}
1&1&\cdots&1
\end{pmatrix}$ is a $\left( 1\times d\right)  $ vector. Vectors $H_t:= \begin{pmatrix}
H_{1,t}& \cdots & H_{k,t} & \cdots & H_{d,t}
\end{pmatrix}^\top$, $B_t := \begin{pmatrix}
B_{1,t}& \cdots & B_{k,t} & \cdots & B_{d,t}
\end{pmatrix}^\top $ and $S_t := \begin{pmatrix}
S_{1,t}& \cdots & S_{k,t} & \cdots & S_{d,t}
\end{pmatrix}^\top$ of dimension $\left( d\times 1\right)  $ each, are, respectively, amount of asset  hold, bought and sold at each decision time $t\in \mathcal{T}_0$. In fact, \eqref{asset_dynamic_1} is obtained by substracting transaction costs in the first equality of \eqref{asset_dynamic}. At time $T$, no more asset is bought or sold: $B_T=S_T=0$; the value of the portfolio is determined by adding all values of assets including the last period returns and external flows. This justifies why there is no transaction cost in \eqref{asset_dynamic_1_}. The reader should notice that variables $cr_t$, $Z_t$, $B_{t}$, $S_{t}$ and $H_{t}$ are all decision variables. We denote by $\bar{H}_k$, the initial holding in asset $k, \; k\in\mathcal{K}$ and $\bar{H}:= \begin{pmatrix}
\bar{H}_{1}& \cdots & \bar{H}_{k} & \cdots & \bar{H}_{d}
\end{pmatrix}^\top $ is a $d \times 1$ vector. $\bar{C}_0$ is the initial cash amount. The first stage asset allocation is determined by
\begin{equation}
H_{0}=\bar{H}+B_{0}-S_{0}
\label{holding_balance_0}
\nonumber
\end{equation}
with total asset
\begin{equation}
A_{0}=\mathbf{e}\cdot \bar{H} + \bar{C}_0 + Z_0 - \left( \bar{c}^B B_{0}+\bar{c}^S S_{0}\right)=\bar{A}_0 +Z_0-  \left( \bar{c}^B B_{0}+\bar{c}^S S_{0}\right)=\mathbf{e}\cdot H_{0}+C_0.
\label{Initial_asset_dynamic_1}
\nonumber
\end{equation}
For $t\geq1 $,
\begin{equation}
H_{t}=\xi_{t}H_{t-1}+B_{t}-S_{t}
\label{holding_balance_1}
\nonumber
\end{equation}
defines the dynamic of holding assets between two consecutive decision times. For any given $\left( k,t\right)$, whenever $S_{k,t}>0, B_{k,t}=0$ and vice-versa. Transaction costs also influence the cash dynamics. Buying an amount $x_k$ of asset $k$ requires $x_k\left(1+\bar{c}_k^B \right)$ of cash and selling the same amount of asset $k$ results in $x_k\left(1-\bar{c}_k^S \right)$ of cash. Initially,
\begin{equation}
C_0=\bar{C}_0 + Z_0 -  \left(\mathbf{e}+\bar{c}^B \right)B_{0} + \left(\mathbf{e}-\bar{c}^S \right) S_{0}
\nonumber
\end{equation}
and for $t\geq 1$,
\begin{equation}
C_t=C_{t-1}\left(1 +r_f \right) + cr_{t-1}W_t - \text{Ben}_t + Z_t - \left(\mathbf{e} + \bar{c}_k^B \right) B_t + \left(\mathbf{e}-\bar{c}_k^S \right) S_t
\nonumber
\end{equation}
where we assume that $cr_{t-1}W_t$, $\text{Ben}_t$ and $Z_t$ come in cash.

\subsubsection{Liability and external flows} \label{Liability}
As we consider a DB plan, total liabilities is the discounted expected value of future pre-defined payments. At a given time $t$, it represents the amount the fund has to reimburse if it has to close at that time. Its value has to be estimated with appropriate rules taking into account actuarial risks, pension fund provisions, and other relevant factors for the employer's line of business. Let $L_t$ denote the total amount of liabilities at time $t$.
\\All quantitative models considered in this paper will be applied to the planning problem of a large and stable pension fund. We can then assume that the fund keeps the same structure and number of members over the study period. Liability, contributions and benefits are therefore invariant with respect to actuarial risk over the period under study. Actuarial risks regroup the random events that affect the number of members into the fund. However, those variables are yearly indexed with the general increase of wages $w_t$. For $t \in \mathcal{T}_1$, we have:
\begin{equation}
L_{t}=L_{t-1} \left(1+ w_{t}\right); \; W_{t}=W_{t-1}\left( 1+w_{t}\right)\; \text{ and } \text{Ben}_t=\text{Ben}_{t-1}\left( 1+\kappa w_{t}\right) 
\end{equation}
and their initial values $L_0$, $W_0$ and $\text{Ben}_0$ known at $t=0$; $\kappa\geq0$ is a model parameter. In practice, the pension payments $\text{Ben}_t$ are often indexed with a certain rate which is a function of the inflation rate.  In order to reduce the complexity of our model, we assume that this indexation rate is a certain proportion of the salary increase as this latter is highly positively correlated to the inflation. From the above definitions, uncertainty, represented by vector $\begin{pmatrix}  \left( 1 + w_t\right), & {\xi_{t} } \end{pmatrix}_{t=1}^T$, affects boths assets and liabilities. As often in the literature (e.g. Kouwenberg \cite{kouwenberg2001scenario}), we use a vector autoregressive model (VAR model) such that:
\begin{equation}
\begin{split}
h_t &= c+\Omega h_{t-1} + \epsilon_t,\;\; \epsilon_t \sim N\left(0,\Sigma \right), 
\\ h_t : &= \begin{pmatrix}
\ln\left( 1+w_t\right) & \ln\left(\xi_{1t} \right) & \cdots & \ln\left(\xi_{kt}\right)  &\cdots&\ln\left(\xi_{dt}\right) 
\end{pmatrix}^\top ,
\\ t & \in \mathcal{T}_1
\end{split}
\end{equation}
where $h_t$ is a $\left\lbrace (d+1) \times 1\right\rbrace $ vector of continuously compounded rate, $c$ the $\left\lbrace (d+1) \times 1\right\rbrace $ vector of coefficients, $\Omega$ the $\left\lbrace (d+1) \times (d+1)\right\rbrace $ matrix of coefficients, $\epsilon_t$ the $\left\lbrace (d+1) \times 1\right\rbrace $ vector of error term and $\Sigma$ the $\left\lbrace (d+1) \times (d+1)\right\rbrace $ covariance matrix. The parameter estimation of this model requires time series analysis. For example in Kouwenberg \cite{kouwenberg2001scenario}, annual observations of the total asset returns and the general wage increase from 1956 to 1994 are used to estimate the coefficients of the VAR model. The resulting estimates will serve in constructing the scenario tree which constitutes the workhorse of multistage stochastic programs.

\subsection{The ALM problem}\label{The ALM problem}
The total cost of funding is the sum of regular ($\sum cr_{t-1}W_t$) and remedial ($\sum Z_t$) contributions over the studied period. In this study, we are looking for the investment stragtegy $H_t$, contribution rate ${cr}_t$ and remedial contribution $Z_t$ for which the total expected cost of funding is minimized. The optimization is made under risk, legal, budget, regulatory and operating constraints. The constraints and objective of the ALM study will be presented in this section.
\\We denote by symbol $\mathbb{E}_{t}\left( x\right)$ the conditional expectation of random variable $x$ with respect to the natural filtration $ \mathcal{F}_t$ whereas $\mathrm {P}\left\lbrace E\right\rbrace $ denotes the occurence probability of event $E$. At each decision time $t$, the optimization problem consists in minimizing the total expected costs under the constraints considered in the following subsections. To simplify the notation, we omit the scenario index $s$.

\subsubsection{Risk constraints}\label{RiskConstraint}
The pension fund wants to guarantee the participants a certain amount of pension. But the members also depend on the pension fund to actually provide for their needs in the future. Therefore, the safety of the portfolio is of paramount concern. This safety is translated in risk constraints.
\\A pension fund has long term obligations, up to decades, and therefore, its planning horizon is large, too. The main goal of an ALM is to find acceptable allocations which guarantee the solvency of the fund during the planning horizon. In general, solvency is measured by the funding ratio ${F}_t$ (also called cover ratio) that we define for a given time $t$ by 
\begin{equation}
{F}_t:=\frac{A_t^*}{L_t}.
\nonumber
\end{equation}
Underfunding occurs when the funding ratio is less than one. The assertion ${F}_t\leq1$ is equivalent to saying that the surplus at time $t$, i.e. $A_t^*-L_t$, is negative. When this occurs, the shortfall could be provided by the fund's sponsor or any other external contribution. That is the remedial contribution as in Haneveld and al. \cite{haneveld2010alm}. Depending on how the random vector $\omega_t:=\begin{pmatrix}  \left( 1 + w_t\right), & {\xi_{t} } \end{pmatrix}, \;t\in\mathcal{T}_1$, behaves, $F_t$ may change over time. Therefore, the pension fund rebalances its assets portfolio and redefines its contribution rate in order to control the funding ratio. The higher $F_t$ is, the healthier the fund is. However, the decision maker would like to avoid as much as possible the changes in contribution rates. We will see in the model description that the parameters can be set in order to limit those variations.
\\The long term objective of the pension fund consists in fullfilling both long and short (one year) term constraints. We define two types of funding ratio risk constraints in this paper. Their goal is to constrain the funding ratio to be larger, on average, than a predefined minimum $\gamma, \gamma\geq0$. Namely, the expected shortfall $\mathbb{E}_{h-1}\left( A_h^*-\gamma L_h  \right)^{-},\; h>t,$ is required to be less than a certain amount $\beta_t$ known at time $t$. Here, $\left( a\right)^-:=\max\left\lbrace -a,0 \right\rbrace $ is the negative part of $a\in\mathbb{R}$. Also in order to simplify understanding, the expression expected shortfall is used to name $\mathbb{E}_{h-1}\left( A_h^*-\gamma L_h  \right)^{-}$. That is slightly different from its definition in actuarial science where $\gamma$ has to be equal to one. The one period risk constraint (OICC\footnote{OICC (resp. MICC) stands for One period Integrated Chance Constraint (resp. Multiperiod Integrated Chance Constraint) which will be more clearly defined in section \ref{framework_ICC}.}) is expressed by
\begin{equation}
\mathbb{E}_{t}\left( A_{t+1}^* - \gamma L_{t+1}  \right)^{-} \leq \beta_t, \;\;t\in \mathcal{T}_0 \setminus \left\lbrace T\right\rbrace.
\label{chance}
\end{equation}
and for the multiperiod (MICC) approach,
\begin{equation}
\mathbb{E}_{h-1}\left( A_{h}^*- \gamma L_{h} \right)^{-}\leq \beta_t, \;\;h\in \mathcal{T}_{t+1}\; \text{and}    \;\; t\in \mathcal{T}_0 \setminus \left\lbrace T\right\rbrace 
\label{chance_1}
\end{equation}
where $\beta_t$ and $\gamma$ are parameters defined by the pension fund. According to the short term approach \eqref{chance}, at each decision time $t$, the expected shortfall over the following period should be smaller than a certain amount $\beta_t$. Notice that, at time $t$, the short term risk only controls the expected shortfall of $\left(  A_{t+1}^* - \gamma L_{t+1} \right)^{-}$ over the following one-year period.
\\When we want to control the expected shortfall over the whole remaining period up to maturity, the risk constraint \eqref{chance_1} is a good measure of long term risk (multiperiod). That is, at time $t$, equation \eqref{chance_1} means that the one period expected shortfall $\mathbb{E}_{h-1}\left( A_{h} - \gamma L_{h} \right)^{-}$ should be smaller than $\beta_t$ at any future node with $h\in \mathcal{T}_{t+1}  $. Equation \eqref{chance_1} can be rewritten as
\begin{equation}
 \max_{h\in \mathcal{T}_{t+1} } \mathbb{E}_{h-1}\left( A_{h}^* -\gamma L_{h}  \right)^{-} \leq \beta_t, \;\; t\in \mathcal{T}_0 \setminus \left\lbrace T\right\rbrace 
\label{chance_2}
\end{equation}
meaning at time $t$, that, the highest one year expected shortfall, over the remaining periods to maturity, has to be smaller than the amount $\beta_t$. Parameter $\beta_t$ is defined at time $t$ by the pension fund and is a function of available informations at that time; e.g. $\beta_t:=f\left( A_t,L_t \right)= \alpha A_t,\;0\leq\alpha\leq1 $. Readers should notice that when $T=1$, $\eqref{chance}$ is equivalent to $\eqref{chance_1}$.
\\In stochastic programming, constraints such as \eqref{chance} and \eqref{chance_1}, i.e. bounding an expected shortfall, are named \emph{integrated chance constraints} (ICC). They were proposed by Haneveld \cite{klein1986duality} as a quantitative alternative for \emph{chance constraints} (CC). In section \ref{Framework}, both ICC and CC will be discussed more specifically. Successful applications of the ICC in ALM for pension fund can be found in Drijver \cite{drijver2005asset} and Haneveld and al. \cite{haneveld2010alm}. The authors assumed that $\beta_t:=\beta$ is unchanged over the studying period and in their numerical illustrations, the ICC is only applied to the first stage. In such case, one can prove that the OICC and the MICC are equivalent. Instead, we remove that assumption in our work. Therefore, we define:
\begin{equation}
\beta_t:=\alpha L_t
\label{def_beta}
\end{equation}
where $\alpha, \;0\leq\alpha\leq1, $ denotes a scale free risk parameter. Until now, we are not aware of any implementation of the MICC \eqref{chance_1} in a multistage framework. OICC is more relax than MICC. Obviously, the two constraints cannot be implemented in the same model at the same time.  There, comes the other particularity of this paper: we analyze the multiperiod risk constraint and then measure how conservative it is comparing to the one period approach.

\subsubsection{Other constraints}
Risk constraints are important, but institutional and legal rules regarding pension fund operations in general are also relevant. As stated in Pflug and al. \cite{pflug1998dynamic}, institutional and legal rules are designed to restrict the risk of losses which would adversely affect pensionners. Hence, the following restrictions are integrated to the model.
\\Firstly, the fund is not allowed to sell an asset that is not owned. This is the \emph{not short selling assets} constraints and can be expressed by
\begin{align}
\nonumber
& \nonumber  H_{k,t}\geq0,\\
&\label{short_selling} B_{k,t}\geq0 \nonumber,\\\nonumber
& S_{k,t}\geq0 \;\; \text{for}\;\; k\in \mathcal{K} \;\; \text{and} \;\; t\in \mathcal{T}_0.
\nonumber
\end{align}
The not short selling constraint goes with the \emph{not borrowing cash} constraint expressed by
\begin{equation}
C_t\geq0,\;\; t\in \mathcal{T}_0.
\nonumber
\end{equation}
Secondly, at any time, the fund should dispose a minimum amount in cash in order to pay eventual claims such as death benefits or pensions. This can be called \emph{liquidity constraint} and is formulated in our model as
\begin{equation}
C_t \left( 1+r_f \right) + \mathbb{E}_{t}\left( cr_tW_{t+1} - Ben_{t+1} \right) \geq0
\nonumber
\label{liquidity}
\end{equation}
which means that, on average, the cash allocation $C_t$ at time $t$ should be sufficient to cover the eventual negative value of the cash flow balance over period $\left[ t, t+1\right]$. Notice that the term \emph{liquidity constraint} used here may have a different meaning in another context, e.g Fonseca and al. \cite{fonseca2007richesse} in a macroeconomic framework.
\\Thirdly, the fund is subject to \emph{portfolio constraint} imposed by the legislator in order to keep a minimum control on its risk exposure. It consists on bounding the holding in asset $k$ by setting upper and lower bounds, $u_k$ and $l_k$ respectively, on $H_{k,t}$. That is
\begin{equation}
l_kA_t  \leq H_{k,t}\leq u_kA_t , \;\; k\in \mathcal{K},\; t\in \mathcal{T}_0 \setminus \left\lbrace T \right\rbrace.
\label{portfolio constraint}
\end{equation}
For example in Switzerland\footnote{OPP2 of April 18$^{\text{th}}$, 1984, Art 55-b, (As of January 1$^{\text{st}}$, 2012)}, the amount allocated to stocks should not exceed fifty percent of total wealth. In such case, $l_{stocks}=0$ and $u_{stocks} = 0.5 A_t$ and constraint \eqref{portfolio constraint} is equivalent to
\begin{equation}
0\leq H_{stocks,t} \leq 0.5  A_t, \;\;\; \;\; t\in \mathcal{T}_0 \setminus \left\lbrace T \right\rbrace .
\nonumber
\end{equation}
These bounds are also applicable to cash $C_t$ and we obtain
\begin{equation}
l_c A_t \leq C_t \leq u_c A_t, \;\;\;\;\; t\in \mathcal{T}_0 \setminus \left\lbrace T \right\rbrace.
\label{portfolio constraint_C}
\end{equation}
Notice that, in equations \eqref{portfolio constraint} and \eqref{portfolio constraint_C}, upper and lower bounds can also be time dependent.
\\Finally in an asset allocation problem, dynamics and budget constraints, already defined in section \ref{Dynamics} are inavoidable. If they were left out, the optimization program would be unbounded. The constraints presented in this subsection are common in any ALM stochastic programming implemention; see for e.g. Kusy and al. \cite{kusy1986bank}, Carino and al. \cite{carino1994russell}, Dempster and Consigli \cite{consigli1998dynamic}, Bogentoft and al. \cite{bogentoft2001asset} and Dert \cite{dert1995asset} among others.

\subsubsection{The optimization problem}
The ALM model is a dynamic decision making optimization tool to minimize the total expected cost under risk and operating constraints. Decisions are taken at the beginning of each one-year period. Accordingly, the ALM model is developped as a multiperiod decision problem, for which, we are asked to come up with an optimal asset allocation, contribution rate and remedial contribution at the beginning of each year. Moreover, penalty costs are assigned to the undesirable events: remedial contributions, and yearly absolute variation of contribution rates. All these components together constitute the objective function:
\begin{equation}
\min_{H,cr,Z} \mathbb{E}_{0}\left[  \sum_{t=0}^{T-1} v_{t+1} \left( cr_tW_{t+1} + \lambda_z Z_{t+1}  \right) + \sum_{t=0}^{T-2} v_{t+1}\lambda_{\Delta_{cr}}  {\Delta_{{cr}_t}}  W_{t+1} \right] 
\label{objective}
\end{equation}
where ${\Delta_{{cr}_t}} :=\; \mid cr_{t+1}-cr_t \mid$ is the absolute variation of contribution rate from year $t$ to $t+1$, $v_t$ is the discount factor for a cash flow in year $t$, $\lambda_z$ and $\lambda_{\Delta_c}$ are, respectively, penalty parameters for remedial contribution and absolute variation of contribution rate. The variables $cr_t$ and $\Delta_{cr_t}$ are bounded:
\begin{equation}
{cr}^l \leq {cr}_t \leq {cr}^u \;\; \text{  and  }  \;\; \underline{\Delta_{cr}} \leq \Delta_{{cr}_t} \leq \bar{\Delta}_{cr}, \;\;\;\;\; t\in \mathcal{T}_0 \setminus \left\lbrace T\right\rbrace 
\nonumber
\end{equation}
where ${cr}^l$, $\underline{\Delta_{cr}}$ are the lower bounds and ${cr}^u$, $\bar{\Delta}_{cr}$ the upper bounds of ${cr}_t$, $\Delta_{{cr}_t}$, respectively. The optimal decisions have to lead to a funding ratio greater than a certain minimum $\bar{F}$ (sometimes called target funding ratio) at the end of period of study $T$:
\begin{equation}
F_T=\dfrac{A_T}{L_T}\geq \bar{F}.
\nonumber
\end{equation}
The entire ALM model, with objective and constraints, can be found in Appendix 1. An optimization program such as \eqref{objective} is often referred to as a \emph{here and now} problem. Uncertainty, characterized by $\omega_t=\begin{pmatrix}  \left( 1 + w_t\right), & {\xi_{t} } \end{pmatrix} ,\; t\in\mathcal{T}_1$, is approached by scenarios. Therefore, we define $\tilde{\omega}$ with a finite number $S$ of possible realizations $\tilde{\omega}^s:=\left( \tilde{\omega}_1^s,\cdots,\tilde{\omega}_T^s \right) , s\in\mathcal{S}:=\left\lbrace1,...,S \right\rbrace $, from $t=0$ to $t=T$ with relative probability $p^s$.\\
\\The objective \eqref{objective} is obviously linear as it can be rewritten as a linear combination of decision variables. We can also notice that dynamics and constraints (except risk constraints which have to be rewritten in a linear form for the stochastic program) presented in sections \ref{Dynamics} and \ref{The ALM problem} are all linear in decision variables. If the risk constraints OICC \eqref{chance} and MICC \eqref{chance_1} were written in a linear form, the ALM problem would be a stochastic linear program (SLP), theoretically solvable by any SLP software depending on its size. In the next section, we will show how they can be turned into linear. The books Kall and Mayer \cite{kall2011stochastic}, Shapiro and al. \cite{shapiro2009lectures} and Birge and Louveaux \cite{birge2011introduction} provide good ressources to deal with such problems. When the size is big, resolution may require heuristic methods. Size is big means that number of asset classes is large or/and time horizon is long or/and number of scenarios is large. Decisions variables are $H_{t}$, $B_{t}$, $S_{t}$, $C_t$, $cr_t$ and $Z_{t}$ for $t\in \mathcal{T}_0 $; but only first stage values $H_{0}$, $B_{0}$, $S_{0}$, $C_0$, $cr_0$ and $Z_0$ are crucial to the decision maker, since, almost surely, a true realization of the random data will be different from the set of generated scenarios.
\\Stochastic programming (SP) is getting popular in ALMs. Its advantage lies in its ability to easily incorporate various types of constraints, Zenios and Bertola \cite{zenios2006practical}. It has rooted with the work of Ziemba and al. \cite{kusy1986bank} who showed, based on a 5-year period application to the Vancouver City Saving Credit Union, that SLP is theoretically and operationally superior to a corresponding deterministic linear programming (LP) model. The authors have proved that the effort required for the implementation of ALM and its computational requirements are comparable to those of the deterministic model. Since then, SP in ALMs has been revisited by many other authors, see, for example, Ziemba and Mulvey \cite{ziemba1998worldwide}, Carino and al. \cite{carino1994russell}, Aro and Pennanen \cite{aro2013liability} and Zenios and Bertola \cite{zenios2006practical} among others.
\\By definition, the pension fund risk problem is often a shortfall problem. In such models, the relevant measure of risk for the firm is the expected amount (if any) by which goals are not met, Carino and al. \cite{carino1994russell}. The model considered in this paper has a general DB ALM structure such as explored in Haneveld and al. \cite{haneveld2010alm} and Ziemba and al. \cite{ziemba1998worldwide}. Its main particularity consists in the integration of ICC by the way of OICC \eqref{chance} and MICC \eqref{chance_1}. A successful implementation of constraints \eqref{chance} in ALM for a DB fund can be found in Vlerk and al. \cite{van2003integrated} and Haneveld and al. \cite{haneveld2010alm}. In their works, the optimization problem is solved assuming that the parameter $\beta_t$ is constant: $\beta_t=\beta$. Furthermore, remedial contribution are provided only when funding ratio falls short in two consecutive years. Implementing this latter condition has lead to the use of binary variables. The authors proposed a heuristic solution to the problem.
\\As we will explain in section \ref{framework_ICC}, the parameter $\beta$ is not scale free. A certain value of $\beta$ does not have equivalent meaning for two different pension funds. It can be too low for a certain fund whereas too high for an other one. In addition, the pension fund actual situation should be taken into account. Our paper is an extension of Haneveld and al. \cite{haneveld2010alm}. As a novelty, we assume that the risk parameter $\beta_t$ vary with respect to time $t$ and is defined  as a proportion $\alpha$ of the actual level of liability at time $t$, see equality \eqref{def_beta}. Roughly speaking, on average, the total asset should cover a proportion of magnitude $\left( 1-\alpha\right) $ of liability at any time. In our model, remedial contribution can however be provided at anytime where solvency is in question, avoiding the use of binary variables, and indirectly, the need of heuristics. Penalty parameter $\lambda_z$ punishes the abuse of remedial contributions.
\\The main features of this study turns around the following points:
\begin{itemize}
\item As in Haneveld and al. \cite{haneveld2010alm} where optimal decision is analyzed for different values of their risk parameter $\beta$, we first measure the effect of our risk parameter $\alpha$ on the decisions $H_0$, $cr_0$ and $Z_0$; this with respect to the OICC. In addition, for a fixed value $\alpha$, the influence of the initial funding ratio is also explored.
\item Secondly, as a safer alternative to the OICC, we propose the MICC (constraint \eqref{chance_1}) and we then measure how hard it is, comparing to the OICC. In constraint \eqref{chance_1}, index $h$ is a decision time index and we are not aware of any implementation of such constraint in ALM. The OICC considered in the first item is actually extended to a multiperiod risk constraint, reinforcing the long term aspect of the pension fund's ALM.
\end{itemize}
In the rest of this paper, OICC (resp. MICC) will stand for the one period (resp. multiperiod) ICC itself as well as the ALM model with the OICC (resp. MICC).

\section{Framework of the risk constraints}\label{Framework}
The most important constraints, of course, deal with the goal of the pension fund: in all circumstances keep a certain control on the funding ratio. This latter is expressed in terms of shortfall constraints which are of ICC type in this paper. Proposed by Haneveld \cite{klein1986duality}, the ICC's formulation direclty results from CC's. That is why, in this chapter, we firstly introduce CC and how it leads to ICC. Secondly, ICC is discussed and we show how constraints $\eqref{chance}$ and $\eqref{chance_1}$ are related to it. We finish this section by proposing simple linear reformulations of $\eqref{chance}$ and $\eqref{chance_1}$.
\\For the sake of clarity, we define the generic linear function $ G:\mathbb{R}^{d}\times\Xi \rightarrow\mathbb{R}^{m} $ such that
\begin{equation}
G\left( X,\omega \right) := BX-D
\nonumber
\end{equation}
where $X\in \mathcal{X}$ is an $d$-vector of decision variables, $\mathcal{X} \subset\mathbb{R}^{d}$ is a polyhedral and closed set and $\omega:=\left( B,D\right)  :\Omega\rightarrow \mathbb{R}^{m} \times \mathbb{R}^{d} \times \mathbb{R}^{m}$ is a random parameter on the probability space $ \left(\Omega,\mathcal{F},\mathrm {P}\right)   $. The support of $\omega$ is defined as the smallest closed set $\Xi\subset \mathbb{R}^{m} \times \mathbb{R}^{d} \times \mathbb{R}^{m}$ having the property $ \mathrm {P}\left(\omega\in\Xi\right)=1$. For $ i \in \mathcal{I}:=\left\lbrace 1,\cdots,m \right\rbrace $, the vector $B$ is of dimension $ \mathbb{R}^{m} \times \mathbb{R}^{d} $ such as $ B:= \begin{pmatrix}
B_1 & \cdots B_i &\cdots & B_m
\end{pmatrix}^\top$ with $B_i \in \mathbb{R}^{d}$ whereas $D := \begin{pmatrix}
D_1 & \cdots D_i &\cdots & D_m
\end{pmatrix}^\top$ with $D_i \in \mathbb{R}$. As supposed in our SP model, we assume that $\omega = \left( B,D\right) $ has a finite number $S$ of possible realizations $\omega^s=\left( B^s,D^s \right) , s\in\mathcal{S}=\left\lbrace1,...,S \right\rbrace $ with respective probability $p^s$.

\subsection{Chance constraints}\label{sec:nothing}
Chance constraints (CC) models serve as tool for modeling risk and risk aversion in SPs. Let $\mathbf{0}$ be a $m$-dimensional vector of zeroes. Satisfying the constraint $G\left(X,\omega  \right) \geq \mathbf{0} $ could lead to high costs or unfeasibilities. This equation refers to a finite system of $m$ inequalities. Instead, if the distribution of $\omega$ is known, one can formulate the condition that the probability of $G\left(X,\omega  \right) \geq \mathbf{0}$ is sufficiently high, i.e. close enough to 1. That is
\begin{equation}
\mathrm {P}\left\lbrace G\left(X,\omega  \right) \geq \mathbf{0}\right\rbrace\geq1-\epsilon
\label{CP}
\end{equation}
where the fixed parameter $\left(1-\epsilon\right) \in \left[ 0,1\right]$ is called probability level and is chosen by the decision maker in order to model the safety requirements. Equation \eqref{CP} is the general form of chance (probabilistic) constraints and can be viewed as a compromise with the requirement of enforcing the constraint $G\left(X,\omega  \right) \geq \mathbf{0}$ for all values $\omega\in\Xi$ of the uncertain data matrix.
\\When $m=1$, $G\left(X,\omega  \right):=g\left(X,\omega  \right) $ is a scalar and equation $\eqref{CP}$ leads to 
\begin{equation}
\mathrm {P}\left\lbrace g\left(X,\omega \right)\geq0\right\rbrace\geq1-\epsilon
\label{ICP}
\end{equation}
with $g:\mathbb{R}^{d}\times\Xi\rightarrow\mathbb{R}$. Equation $\eqref{ICP}$ is known as \emph{individual} CC. For $m>1$, we obtain
\begin{equation}
\mathrm {P}\left\lbrace g_i\left(X,\omega \right)\geq0,\; \; i\in\mathcal{I}\right\rbrace\geq1-\epsilon,
\label{JCP}
\end{equation}
called \emph{joint} CC.
\\Chance-constrained programs have been pionnered by Charnes and al. \cite{charnes1958cost} in production planning. Since then, they have been extensively studied and have also been applied in many other areas such as telecommunication, finance, chemical processing and water ressources management. Despite important theoretical progress and practical importance, there could be major problems with numerical processing of CCs, see Ahmed and Shapiro \cite{ahmed2008solving} and Nemirovski and Shapiro \cite{nemirovski2006convex}.
\\Especially when $\omega $ has a discrete distribution, Raike \cite{raike1970dissection} introduces a mixed-integer reformulation of CC.
Assuming $m=1$, equation $\eqref{ICP}$ is equivalent to
\begin{equation}
\sum_{s=1}^{S} p^{s}\cdot\mathbf{1}_{\left(g\left(X,\omega^s\right)\geq0\right)}{\left(s\right)}\geq1-\epsilon
\nonumber
\end{equation}
where $\mathbf{1}_{\left(g\left(X,\omega^s\right)\geq0\right)}{\left(s\right)}=1$ if $g\left(X,\omega^s\right)\geq0$ and $0$ otherwise. Now, we are able to write inequalities $\eqref{ICP}$ in a mixed-integer program (MIP) formulation. We introduce binary variables $\delta^s,\;s\in\mathcal{S}$. They play the role of indicator function: $\delta^s=1$ in scenario $s$ if it holds that $g\left(X,\omega^s\right)<0$ and equals $0$ otherwise. In terms of these additional decision variables, the CC can be written as linear inequalities
\begin{eqnarray}
g^s\left(X,\omega^s\right)+\delta^sM\geq 0,\;\;s\in\mathcal{S},\label{binaryCP1}
\\\sum_{s=1}^{S} p^{s}\delta^s\leq\epsilon,\;\;s\in\mathcal{S},\label{binaryCP2}
\\x\in{X},\;\;\delta^s\in\left\lbrace0,1 \right\rbrace,\;\;s\in\mathcal{S},\label{binaryCP3}
\end{eqnarray}
where $M$ is a sufficiently large number. If $\delta^s=0$, then the constraint $g\left(X,\omega^s\right)\geq0$ corresponding to the realization $s$ in the sample is enforced. On the other hand, if $\delta^s=1$, the constraint is satisfied for any candidate solution. The probability weighted average of these binary variables equals the risk of not meeting the condition $g\left(X,\omega^s\right)\leq0$ with the decision $X$, which should be at most $\epsilon$.
\\This formulation is well known in SP and has first been applied to ALM for pension fund by Dert \cite{dert1995asset}. It also holds for the joint CC case where $m>1$. In fact, $\left\lbrace g_i\left(X,\omega\right)\geq0,\; \; i\in\mathit{I}\right\rbrace$ is equivalent to
\begin{equation}
\min_{i\in\mathit{I}} \left\lbrace g_i\left(X,\omega\right)\right\rbrace\geq0
\nonumber
\end{equation}
and can also be writen as linear inequalities
\begin{eqnarray}
g_i^s\left(X,\omega^s\right)+\delta_i^sM\geq 0,\;\; i\in\mathit{I},\;\;s\in\mathcal{S}, \label{binaryJCP1}
\\ \Delta^s\geq\delta_i^s, \;\;\;\;\;\; i\in\mathit{I},\;\;s\in\mathcal{S}, \label{binaryJCP4}
\\\sum_{s=1}^{S} p^{s}\Delta^s\leq\epsilon, \;\;\;\;\;\;\;\;\;\;\;\;\;\;\;\;\;\;\;s\in\mathcal{S}, \label{binaryJCP2}
\\ x\in{X},\;\;\Delta^s,\delta_i^s\in\left\lbrace0,1 \right\rbrace,\;\;i\in\mathit{I},\;\;s\in\mathcal{S}.\label{binaryJCP3}
\end{eqnarray}
Even with these linear settings $\eqref{binaryCP1}-\eqref{binaryCP3}$ and $\eqref{binaryJCP1}-\eqref{binaryJCP3}$, implementing this constraint with a reasonnable number of scenarios can be computationally challenging as the feasible set is obviously not linear, neither convex. That is due to the increase in complexity from MIP that arises from the introduction of at least one binary variable per each of the $S$ scenarios. Efficient solution algorithms are proposed in chapter $4$ of Kall and Mayer \cite{kall2011stochastic}, Luedtke \cite{luedtke2013branch}, Luedtke and al. \cite{luedtke2010integer}, Tanner and Ntaimo \cite{tanner2010iis}, Prekopa and al. \cite{prekopa1998programming} and Ruszczynski \cite{ruszczynski2002probabilistic}.
\\Note that the CC, as described above, only considers the qualitative aspect of the risk, i.e. attention is only paid to whether the integrand is satisfied or not. A better approach can be to control the quantitative aspect of the fail, i.e the size of negative values of $G^s$. That is often the case for pension funds where sponsors want to know approximatively how much they are willing to contribute in the following periods. Due to an idea of Haneveld \cite{klein1986duality}, binaries $\delta^s$ are dropped and the integrated chance constraint has been proposed.

\subsection{Integrated Chance Constraint}
\label{framework_ICC}
The MIP constraints $\eqref{binaryJCP1}$ to $\eqref{binaryJCP3}$ are hardly implementable due to the integrality conditions in $\eqref{binaryJCP3}$. For problems involving binary (or general integer) decision variables, a natural approach is to relax the integrality and solve the resulting relaxation, see Vlerk and al. \cite{van2003integrated}. If we relax the integrality constraints and substitute $y^s:=\delta^sM$ and $\beta:=\alpha M$, we obtain
\begin{eqnarray}
B^sX+y^s\geq{D^s},\;\;s\in\mathcal{S}\label{ICC1}
\\\sum_{s=1}^{S} p^{s}y^s\leq\beta,\;\;\;\;\;\;\;\;\;\;\;\;\label{ICC2}
\\y^s\geq0,\;\;\;\;s\in\mathcal{S}\label{ICC3}
\\X\in{\mathcal{X}},\;\;\;\;\;\;\;\;\;\;\;\; \label{ICC5}
\end{eqnarray}
where the parameter $\beta$ is non-negative. By \eqref{ICC1}, for each $s$, the non-negative variable $y^s$ is not less than the shortfall $\left( B^sx-D^s\right)^-$, where $\left( a\right)^-:=\max\left\lbrace -a,0 \right\rbrace $ is the negative part of $a\in\mathbb{R}$. The inequality $\eqref{ICC2}$ therefore puts an upper bound $\beta$ on the expected shortfall. That is, the system $\eqref{ICC1}-\eqref{ICC5}$ is equivalent to
\begin{equation}
\mathbb{E}\left(BX-D\right)^{-}=\sum_{s=1}^{S} p^{s}\left( B^sX-D^s\right)^- \leq\beta.\label{ICC_expect}
\end{equation}
Such constraint is called integrated chance constraint (ICC) and has been introduced by Haneveld \cite{klein1986duality} as an alternative to CC. However, Haneveld and al. \cite{haneveld2010alm}, Vlerk and al. \cite{van2003integrated} and Drijver \cite{drijver2005asset} have pionnered its application to ALM for pension funds and since then, it has been implemented in practice.
\\By definition, the feasible set, defined by linear inequalities $\eqref{ICC1}-\eqref{ICC5}$ is a polyedron (convex) as it contains only continuous decision variables, see Haneveld and Vlerk \cite{haneveld2002integrated}. Thus, it can usually be solved efficiently using an appropriate software. Constraints $\eqref{ICC1}-\eqref{ICC5}$ are very attractive from an algorithm point of view. Haneveld and Vlerk \cite{haneveld2002integrated} propose a faster algorithm for big size problems. ICC is a good alternative to CC from different perspectives:
\begin{itemize}
\item Firstly, CC only measures the probability of shortage whereas ICC uses the probability distribution to measure the expected magnitude of the shortage. We can say that ICC takes into account both quantitative and qualitative aspects of the shortage whereas CC only considers its qualitative side. CC says only if there is underfunding or not and especially in practice, it could be important to limit the amount of remedial contributions the sponsor is willing to provide in years after.
\item Secondly, ICC and CC somehow ressemble, respectively, to the so-called \emph{conditional value-at-risk} (CVaR) and \emph{value-at-risk} (VaR). Conversely to CVaR which is known as \emph{coherent} (Rockafellar and Uryasev \cite{rockafellar2002conditional}), it is well known  that VaR is not a coherent risk measure as it does not fulfill the subadditivity condition. Therefore, ICC possesses more attractive risk properties than CC. To learn more about coherent risk measures, see Artzner and al. \cite{artzner1999coherent}.
\item We should also add that, if the risk aversion parameter is changed, the feasible region in case of ICC changes smoothly, while this region changes in a rough way in case of CC, Drijver \cite{drijver2005asset}.
\item Finally, we should admit that the parameter $\epsilon$ of CCs is scale free, and corresponds to risk notion which is more familiar to pension fund managers. It is not the case for ICC. Our solution to this problem is to set $\beta$ as a proportion $\alpha$ of liability.
\end{itemize}
From now, and without loss of generality, we assume $m>1$. Therefore, equation \eqref{ICC_expect} can be rewritten as
\begin{equation}
\mathbb{E} \left\lbrace \left(B_i X-D_i \right)^{-},\;i\in\mathit{I}\right\rbrace  \leq\beta\label{ICC_expect_1}
\nonumber
\end{equation}
which is the \emph{joint} form of ICC, see Haneveld and Vlerk \cite{haneveld2002integrated}. When index $i$ is a decision stage index with conditional expectation at stage $i$, we obtain a multistage program and variable $X$ becomes stage dependent ($X_i$). That is, at stage $j\in \mathit{I} \setminus \left\lbrace m \right\rbrace $:
\begin{equation}
 \mathbb{E}_{i} \left(B_{i+1} X_{i}-D_{i+1} \right)^{-} \leq\beta_j,\; \; i\in \left\lbrace j,j+1,\cdots,m-1\right\rbrace  \label{ICC_expect_2}
\end{equation}
which is equivalent to the MICC $\eqref{chance_1}$ for $\mathit{I}=\mathcal{T}_0 $ and $B_{h+1}X_h-D_{h+1}=A_{h+1}-\gamma L_{h+1}$. At time $t$, that is:
\begin{equation}
 \mathbb{E}_h  \left(A^{*}_{h+1}-\gamma L_{h+1} \right)^{-}\leq\beta_t,\;h\in \mathcal{T}_t \setminus \left\lbrace T\right\rbrace .\label{ICC_expect_3}
\end{equation}
The parameter $\beta_t$ is then set at time $t$ and will remain applicable until $T$. As decision is taken at each stage, the MICC inequality $\eqref{chance_1}$ shows a collection of inequality \eqref{ICC_expect_3} going from $t=0$ to $t=T-1$. Similarly, when $m=1$, one can proove that equation $\eqref{ICC_expect}$ leads to the OICC $\eqref{chance}$.

\subsection{OICC and MICC: Scenario tree interpretation} \label{ICC_interpretation}
Section \ref{multistage recourse models} briefly explains our scenario tree model. We recall that the node $\left( t,s\right) $ corresponds to a certain scenario $s$ at decision time $t$. To avoid anticipativity, we have to consider that many nodes $\left( t,s\right) $ might correspond graphically to the same thing on the scenario tree picture. For example in Figure \ref{arbre_scenario}, the nodes $\left( 1,1\right),\left( 1,2\right),\cdots,\left( 1,8\right) $ correspond graphically to the empty red cercle. At each node $\left( t,s \right) $, the fund's manager has to rebalance the asset portfolio and fix the contribution rate. These decisions are taken considering the actual scenario and possible future paths as well as the risk constraints.

\subsubsection{OICC}
In principle, considering a certain node $\left( t,s \right) $, the OICC constraint $\eqref{chance}$ would be implemented as follows:
\begin{equation}
\mathbb{E}_{t,s}\left( A_{t+1}^{*s} - \gamma L_{t+1}^s\right)^{-} := \sum_{s^{'} \in \mathcal{S}} p_{t,s}^{s^{'}} \left( A_{t+1}^{*s^{'}} - \gamma L_{t+1}^{s^{'}} \right)^{-} \leq \alpha L_{t}^s
\label{chance_prime}
\end{equation}
where $p_{t,s}^{s^{'}}$ stands for the conditional probability to reach node $\left(t+1,s^{'}\right)$ going from $\left( t,s\right)$ and $p_{t,s}^{s^{'}}=0$ for any scenario $s^{'}$ of $t+1$ not descending from $\left( t,s\right) $. As in Vlerk and al. \cite{van2003integrated}, we include the linear inequality \eqref{chance_prime} in every subproblem $\left( t,s\right),\; t<T $ of our multistage recourse model. At $\left( t,s\right) $, they reflect the short-term risk constraint, stating that the expected funding shortfall over the following period $\left( t+1\right) $ is at most $\alpha L_{t}^s$. In other words, on average, the pension fund should be able to cover the proportion $\left( 1-\alpha \right) $ of its total liability. The optimization problem will be more and more relax with the increase in $\alpha$.

\subsubsection{MICC}
Considering the node $\left( t,s\right) $, the MICC constraint can be formulated in the following way:
\begin{equation}
 \mathbb{E}_{h-1,s}\left( A_{h}^{*s} -  \gamma L_{h}^s \right)^{-} \leq \alpha L_t^s, \;\;h\in \mathcal{T}_{t+1} \;  \text{and} \;\; t\in \mathcal{T}_0 \setminus \left\lbrace T\right\rbrace
\label{chance_prime1}
\end{equation}
with
\begin{equation}
\mathbb{E}_{h-1,s}\left( A_{h}^{*s} -  \gamma L_{h}^s \right)^{-}=\sum_{s^{'} \in \mathcal{S}} p_{h-1,s}^{s^{'}} \left( A_{h}^{*s^{'}} - \gamma L_{h}^{s^{'}} \right)^{-}.
\nonumber
\end{equation}
Under \eqref{chance_prime1}, at each node $\left( t,s \right) $, decisions are taken such that the descending nodes's one-period expected shorfall are smaller than $\alpha L_t^s$ (defined at current node). Such constraint permits to have a certain control of the cover ratio over the whole remaining periods: $\left[ t+1,T\right] $; whereas \eqref{chance_prime} only covers one period: $\left[ t,t+1 \right] $. For example, at initial time $t=0$, the minimum cost is defined such that the expected shortfall at any node in the tree (as descendant of the initial node) is smaller than $\beta_0=\alpha L_0^s$ as in Haneveld and al. \cite{haneveld2010alm}:

\begin{equation}
\sum_{s^{'} \in \mathcal{S}} p_{t,s}^{s^{'}} \left( A_{t+1}^{*s^{'}} - \gamma L_{t+1}^{s^{'}} \right)^{-}\leq \beta_0 , \;\;\; t\in \mathcal{T}_0\setminus \left\lbrace  T\right\rbrace ,\; s\in \mathcal{S}.
\nonumber
\end{equation}
Futhermore, at each node $\left( t,s\right), t\in \mathcal{T}_1\setminus \left\lbrace  T\right\rbrace , s\in \mathcal{S}$, we add the restriction:
\begin{equation}
\sum_{s^{'} \in \mathcal{S}} p_{t,s}^{s^{'}} \left( A_{t+1}^{*s^{'}} - \gamma L_{t+1}^{s^{'}} \right)^{-}\leq \alpha L_t^s.
\nonumber
\end{equation}
That is how we implement \eqref{chance_prime1} at initial node. If we repeat the same procedure at each node of the tree, we can then propose a simpler SP reformulation:
\begin{prop}
\label{my_proposition}
Constraint \eqref{chance_prime1} is equivalent to the following linear statement:
\\At each node $\left( t,s \right),\; t<T,\; s\in \mathcal{S} $
\begin{equation}
\sum_{s^{'} \in \mathcal{S}} p_{t,s}^{s^{'}} \left( A_{t+1}^{*s^{'}} - \gamma L_{t+1}^{s^{'}} \right)^{-} \leq \min_{0\leq t'\leq t} \alpha L_{t'}^s. 
\label{chance_prime2}
\end{equation}
\end{prop}
That is, at a given node $\left( t,s \right), t<T, s\in \mathcal{S} $, the expected shortfall over the next period should be less or equal to the smallest value of $\alpha L_{t'}^s$ calculated over the preceding nodes $\left( t',s \right), t'\leq t$. This is based on the fact that, in the multiperiod framework, decision taken at node $\left( t,s \right) $ is influenced by the history of $\omega_t^s$ up to time $t$, in particular $\beta_{t}^s$ at preceding nodes. Inequality \eqref{chance_prime2} is linear and describes a polyhedral set. The proof of proposition \ref{my_proposition} is straightforward when we go backward in time starting from nodes $\left( T-1,s \right)$, see Appendix 2 for an example based sketch of proof. At each node $\left( t,s \right) $, as we know the history of $\beta_{t}^s$ up to time $t$, one can determine the smallest $\beta_{t'}^s,\; {t'}\leq t$. Therefore, implementation of MICC consists in including the linear constraint \eqref{chance_prime2} at each node $\left( t,s \right) $.

\section{Numerical illustrations} \label{Num_ill}
This section contains computational results for the SP model. Let's recall that we are dealing with a DB pension fund whose objective is to minimize the total expected costs under constraints. The study will focus on risk constraints which are of ICC type. Firstly, based on the OICC, the effects of risk parameter and cover ratio on the optimal decisions are analyzed. Prima facie, the MICC appears to be a safer and more restrictive than OICC. Based on the same analysis as before, the cost of conservativeness is subsequently measured.
\\For this study, consider a hypothetical large pension fund which may invest into $d=4$ classes of asset ordered by level of risk:
\begin{enumerate}
\item Deposits,
\item Bonds,
\item Real estate,
\item Stocks.
\end{enumerate}
We are aware of the fact that the number of assets is often much larger in practice. That said, only four classes of assets are considered here in order to reduce the complexity of the model. After investing in these asset classes, the rest is held in cash. The deterministic properties of asset classes are described in Table \ref{deterministic_properties}. Investment limits are defined with respect to practical rules of liquidity and diversification; transaction costs are taken from Haneveld \cite{haneveld2010alm} with $\bar{c}^S=\bar{c}^B=\bar{c}$, whereas the initial investments are defined considering general statistics of pension fund's assets allocation in Switzerland, see Towers Watson \cite{watson2013global} (where we assume that "real estate" corresponds to "other assets"). The portfolio constraints are defined in term of proportion and all amounts are assumed to be in thousands of Swiss francs. The values of the other deterministic parameters are shown in Table \ref{deterministic_parameters_values}
\\The time horizon $T=5$ is split into five periods of one year each. Consequently, the considered ALM model has five stages, allowing for decisions at $t=0$ (now) up to time $t=4$. The random vector $\omega_t$ follows a VAR process, approximated in our case by a multistage scenario tree. In the following considerations, we firstly present the descriptive statistics of our model and the numerical results, obtained from our study, are discussed on a second hand.

\begin{table}[t]
\centering
\caption{\label{deterministic_properties} Data on the asset classes }
\begin{tabular}{|l|l|l|l|l|l|} 
   \hline
   \hline
    Asset classes & $k$ & $l_k$ & $u_k$ & $\bar{c}$ & Initial Investments \\
    \hline
    \hline
    Cash & $-$ & $0$ & $1$ & $0$ & 4'950 \\
       Deposits & $1$ & $0$ & $0.5$ & 0.00150 & 16'500 \\ 
       Bonds & $2$ & $0.1$ & $1$ & 0.00150 & 38'500  \\
       Real estate & $3$ & $0$ & $0.30$ & $0.00425$ & 17'600 \\
       Stocks & $4$ & $0$ & $0.50$ & 0.00425 & 32'450 \\
    \hline
    \hline
\end{tabular}
\end {table}

\begin{table}[t]
\centering
\caption{\label{deterministic_parameters_values} Values of the other deterministic parameters}
\begin{tabular}{|l|l|l|l|}
\hline
\hline
$\lambda_Z=350$ \;\;\;\;\;$\lambda_{\Delta_{cr}}=1$\;\;\;\;\;$r_f=0.008$\;\;\;\;\;$v_t=(1+r_f)^{-t}$\\
$\underline{\Delta_{cr}}=-0.08$\;\;\;\;\; ${cr}^l=-0.08$\;\;\;\;\;$\bar{\Delta}_{cr}=0.05$\;\;\;\;\;${cr}^u=0.3$\\
$\bar{A}_0=110'000$\;\;\;\;\;$\gamma=1.05$\;\;\;\;\;$\bar{F}=1.05$\\

\hline
\hline
\end{tabular}
\end{table}

\subsection{Scenarios}
The implementation of the scenario tree requires a careful specification of the VAR process. For this purpose, we use the estimation results obtained in Kouwenberg \cite{kouwenberg2001scenario}. More specifically, the author estimates this process based on annual observations of the total asset returns and the general wage increase from 1956 to 1994. Table \ref{Stat_desc} displays descriptive statistics of the time series whereas Table \ref{erreur_corr} shows the estimated correlation matrix of the residuals. Future returns for financial planning models can be constructed by sampling from the error distribution of the VAR model and applying the estimated equations of Table \ref{parameters}. We refer to Kouwenberg \cite{kouwenberg2001scenario} for further details on this model estimation and for building the tree as well. For this purpose, we specify a branching structure of $1-10-6-6-4-4$. This scenario tree has one initial node at time $0$ and $10$ succeeding nodes at time $1$, $\cdots$, resulting in $10 \times 6 \times 6 \times 4 \times 4=S=5760$ path from $0$ to $5$, each with probability $p^s=\frac{1}{5760}$.

\begin{table}[t]
\centering
\caption{\label{Stat_desc} Statistics, time series 1956-1994, Kouwenberg \cite{kouwenberg2001scenario} }
\begin{tabular}{|l|r|r|r|r|} 
   \hline
     & \multicolumn{4}{c|}{Statistics} \\
    \cline{2-5}
    \cline{2-5}
     Assets & Mean & S.D. & Skewness & Kurtosis \\
    \hline
    \hline
    Wages & $0.061$ & $0.044$ & $0.434$ & $2.169$ \\ 
    Deposits & $0.055$ & $0.025$ & $0.286$ & $2.430$ \\
    Bonds & $0.061$ & $0.063$ & $0.247$ & $3.131$ \\
    Real estate & $0.081$ & $0.112$ & $-0.492$ & $7.027$ \\
    Stocks & $0.102$ & $0.170$ & $0.096$ & $2.492$ \\
    \hline
\end{tabular}
\end {table}
\begin{table}[t]
\centering
\caption{\label{erreur_corr} Residual correlations of VAR-model, Kouwenberg \cite{kouwenberg2001scenario} }
\begin{tabular}{|l|c|c|c|c|c|} 
   \hline
    Assets & Wages & Deposits & Bonds & Real estate & Stocks \\
    \hline
    \hline
    Wages & $1$ &  &  &  & \\ 
       Deposits & $0.227$ & $1$ &  &  & \\
       Bonds & $-0.152$ & $-0.268$ & $1$ &  & \\
       Real estate & $-0.008$ & $-0.179$ & $0.343$ & 1 & \\
       Stocks & $-0.389$ & $-0.516$ & $0.383$ & $0.331$ & $1$\\

    \hline
\end{tabular}
\end {table}
\begin{table}[t]
\centering
\caption{\label{parameters} Coefficient of the VAR model, Kouwenberg \cite{kouwenberg2001scenario} }
\begin{tabular}{|l|} 
   \hline
   \hline
   $\begin{array} {lccclcccl} 
   \ln\left( 1 + \text{wages}_t\right)  & = & 0.018 & + & 0.693 \ln\left( 1 + \text{wages}_{t-1}\right) + e_{1t}   &   &  &   &  \sigma_{1,t}=0.030
   \\   &    & (2.058) &   & (5.789) &   &  &   & 
   \\   &    & &   &  &   &  &   & 
   \\ \ln\left( 1+\text{deposits}_t\right) & = &  0.020 & + & 0.644 \ln\left( 1+\text{deposits}_{t-1}\right) + e_{2t}  &   &  &   & \sigma_{2,t}=0.017
   \\   &    & (2.865) &   & (5.448) &   & &   & 
   \\   &    & &   &   &   & &   & 
   \\ \ln\left( 1+\text{bonds}_t\right) & =  & 0.058 & + & e_{3t}  &   &  &   & \sigma_{3,t}=0.060
   \\   &    & (6.241) &   &  &   & &   & 
   \\   &    &  &  &  &   & &   & 
   \\ \ln\left( 1+\text{real estate}_t\right) & =  & 0.072 & + & e_{5t}  &   &  &   & \sigma_{5,t}=0.112
   \\   &    & (4.146)  &   &  &   &
   \\   &    &   &  &  &   &  &   & 
   \\ \ln\left( 1+\text{stocks}_t\right) & =  & 0.086 & + & e_{6t}  &   &  &   & \sigma_{6,t}=0.159
   \\   &    & (3.454) &   &  &   &  &   & 
   \end{array}$
    \\
    \hline
    \hline
\end{tabular}
\end {table}


\subsection{Numerical results}
This section presents the outputs of our study. All numerical results were implemented using the solver CPLEX in the mathematical programming language AMPL. The ALM models are formulated as  large LP-problems with $616'321$ variables. In the model with the OICC, there are $995'347$ constraints and $3'041'032$ nonzeros in the constraint matrix whereas they are respectively $1'002'317$ and $3'105'602$ in the MICC. On average, the solution times are $381$ seconds and $448$ seconds, respectively, for OICC and MICC.
\\As a result of the ALM analysis, we are supposed to provide the first stage optimal decisions: a contribution rate, a remedial contribution and asset allocation that minimize the total cost. In the first part of this section, we analyze the effects of the risk parameter $\alpha$ and the initial funding ratio $F_0$ on the optimal decision. The optimization is made with respect to OICC. The values of $\alpha$ ranges from $0$ to $0.085$ whereas the initial funding ratio $F_0=\frac{\bar{A}_0}{L_0}$ vary from $0.5$ to $1.5$. In order to vary $F_0$, we change the initial liability $L_0$ accordingly, as $\bar{A}_0$ is specified from Table \ref{deterministic_parameters_values}. In the second part, we compare the OICC to the MICC.

\subsubsection{OICC}
\begin{figure}[h!] 
	\begin{minipage}[c]{.46\linewidth} 
		\includegraphics[width=7.5cm,height=6.5cm]{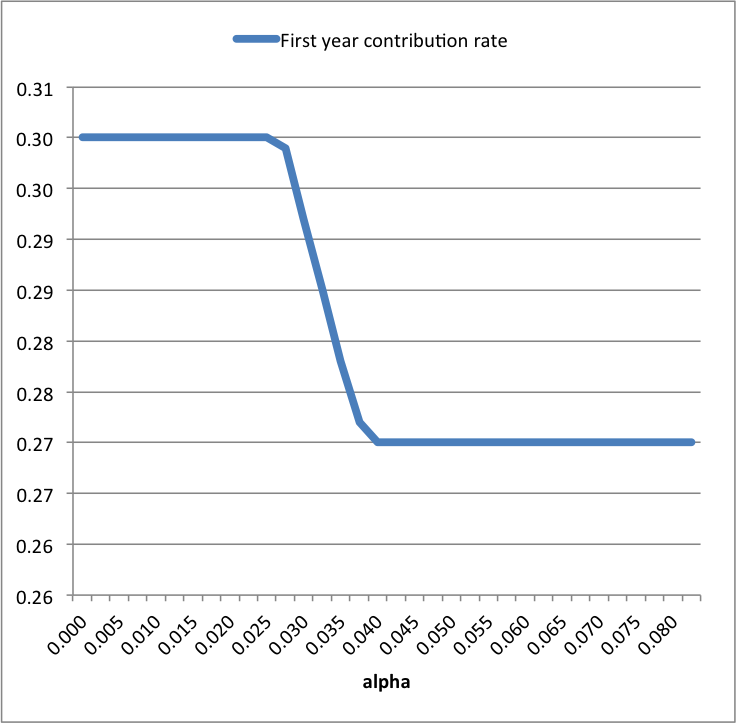}\vspace{0.cm} 
		\caption{OICC: Contribution rate at $t=0$ as function of $\alpha$} 
		\label{OICC_alpha_crate} 
	\end{minipage} \hfill
	\begin{minipage}[c]{.46\linewidth} 
		\includegraphics[width=7.5cm,height=6.5cm]{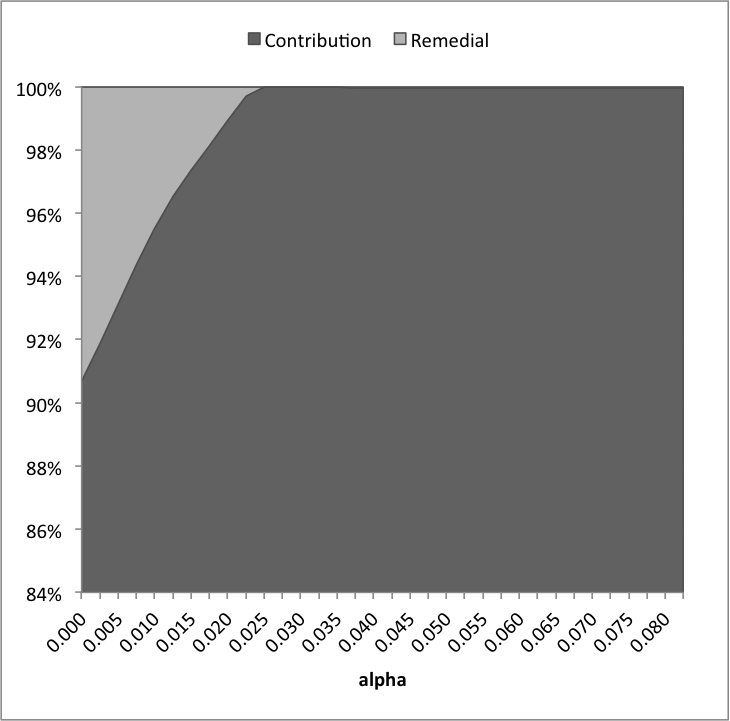}\vspace{0.cm}
		\caption{OICC: Initial cost allocation in function of $\alpha$}
		\label{OICC_alpha_contrem}
	\end{minipage}
\end{figure}
\begin{figure}[h!]
	\begin{center}
		\includegraphics[width=12cm,height=7cm]{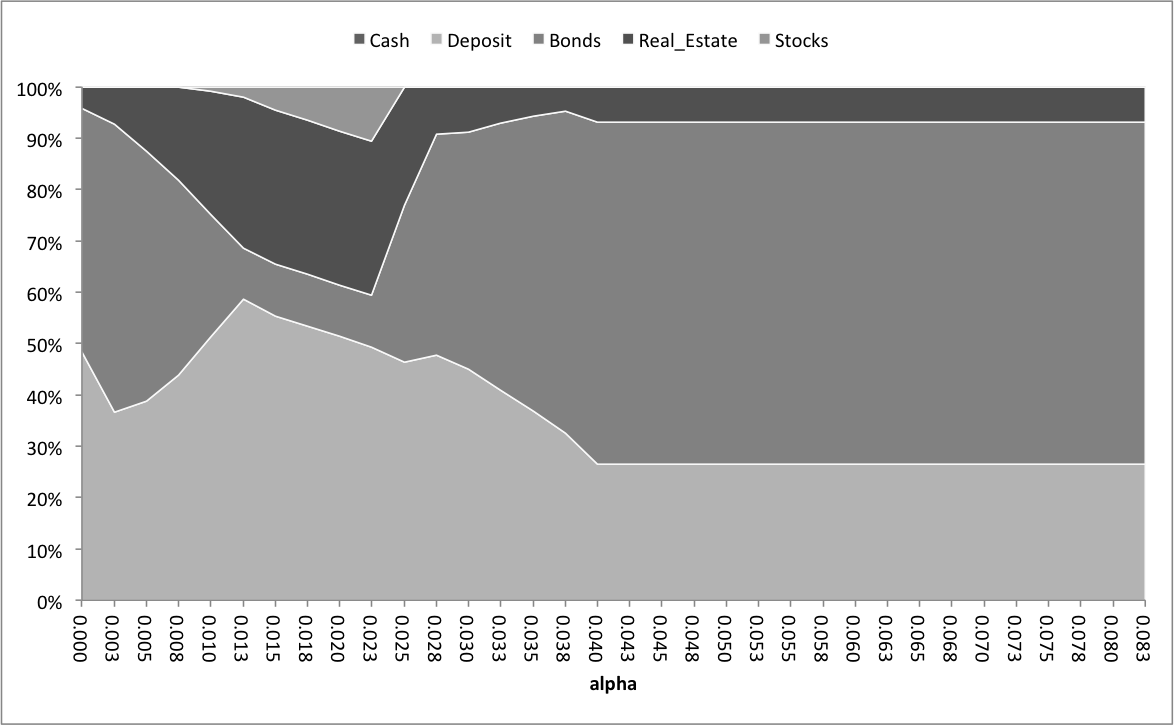} \caption{\label{OICC_alpha_allocation}OICC: Asset allocation at $t=0$ in function of $\alpha$} 
	\end{center}
\end{figure}~\\
\\In order to analyze the impact of the risk parameter $\alpha$, we fix the value of the initial cover ratio. That is:
\begin{equation}
L_0:=120'000\Rightarrow F_0=\dfrac{\bar{A}_0}{L_0}=0.9166, \Rightarrow \underline{\text{under covered}}.
\nonumber
\end{equation}
In what follows, the letter $O$ at the top of a symbol stands for OICC whereas $M$ is related to MICC. Figure \ref{OICC_alpha_crate} shows the evolution of the contribution rate $cr_0$. The value of $cr_0$ is particularly high as the institution is underfunded. We osberve that when $\alpha\leq\alpha^{O}_*:=0.025$ ($O$ at the top stands for OICC.), the contribution rate is at its maximum: ${cr}^u=0.3$ as specified in Table \ref{deterministic_parameters_values}. From $0.025$, ${cr}_0$ decreases linearly until it reaches the value $0.27$ at $\alpha=\bar{\alpha}^O:=0.04$, and remains unchanged thereafter. According to the objective \eqref{objective}, the total cost is composed of the total contribution and of the total remedial contribution, these, over the period under study. Remedial contribution should be seen as an external financial support which may come from the sponsor of the pension fund. Figure \ref{OICC_alpha_contrem} displays the allocation of the total costs into the two types of contribution. The proportion of remedial contribution linealy decreases from $9\%$ at $\alpha=0$ to reach $0\%$ at $\alpha^O_*=0.025$ and stays constant for $\alpha\geq \alpha^O_*$. Indeed, Figures \ref{OICC_alpha_crate} and \ref{OICC_alpha_contrem} help understand how the ALM model parameters have been defined. It is conventional to assume that, from a certain level of risk and for a fix cover ratio, the sponsor will no more provide any financial support to the fund. In this model, the penalty parameter $\lambda_z$ has been set such that the total remedial contribution is zero for $ \alpha \geq \alpha^O_* $. Consequently, $cr_0$ decreases from $\alpha=\alpha^O_*$. It remains equal to $0.27$ for $\alpha\geq \bar{\alpha}^O$ due to the target cover ratio at maturity: $F_5\geq\bar{F}=1.05$.
\\Figure \ref{OICC_alpha_allocation} describes the optimal asset allocation for different values of $\alpha$. Assets are ordered with respect to their level of risk. For small values of the risk parameter, the proportion of riskier assets (stocks and real estate) tends to increase with an increase of $\alpha$. When it approaches $\alpha^O_*$, as the remedial contribution is already low, the decision maker starts reducing the risk exposition of its portfolio. However, the proportion of bonds is increased in order to improve the performance of the asset portfolio. The risk exposition is then progressively reduced until $\alpha=\bar{\alpha}^O$, from which, it remains unchanged thereafter. The value $\bar{\alpha}^O$ can be seen as the smallest value of $\alpha$, above what, the OICC influence is no more significant, i.e. contribution rate, remedial contribution and asset allocation stay constant.
\begin{figure}[h!] 
	\begin{minipage}[c]{.46\linewidth} 
		\includegraphics[width=7.5cm,height=6.5cm]{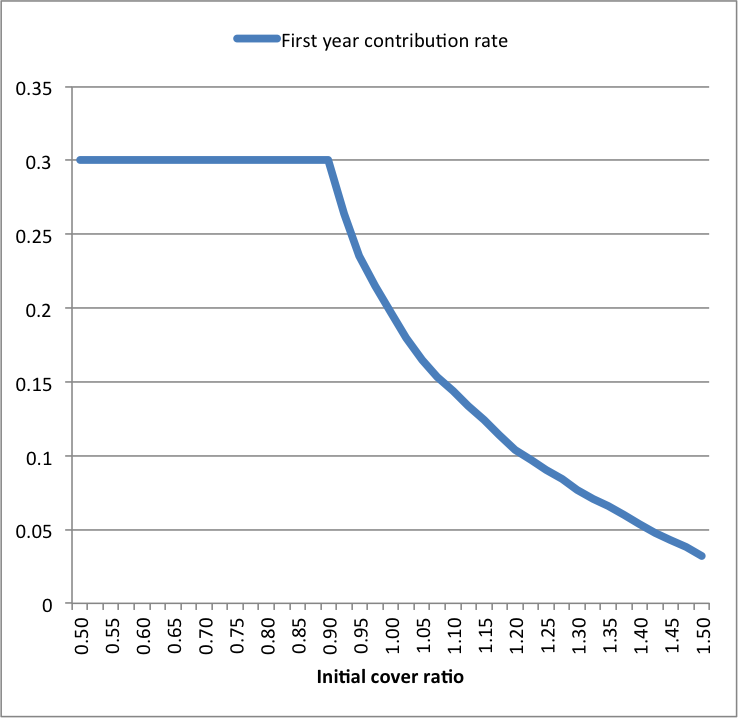}\vspace{0.cm} 
		\caption{OICC: Contribution rate at $t=0$ in function of $F_0$} 
		\label{OICC_CR_crate} 
	\end{minipage} \hfill
	\begin{minipage}[c]{.46\linewidth} 
		\includegraphics[width=7.5cm,height=6.5cm]{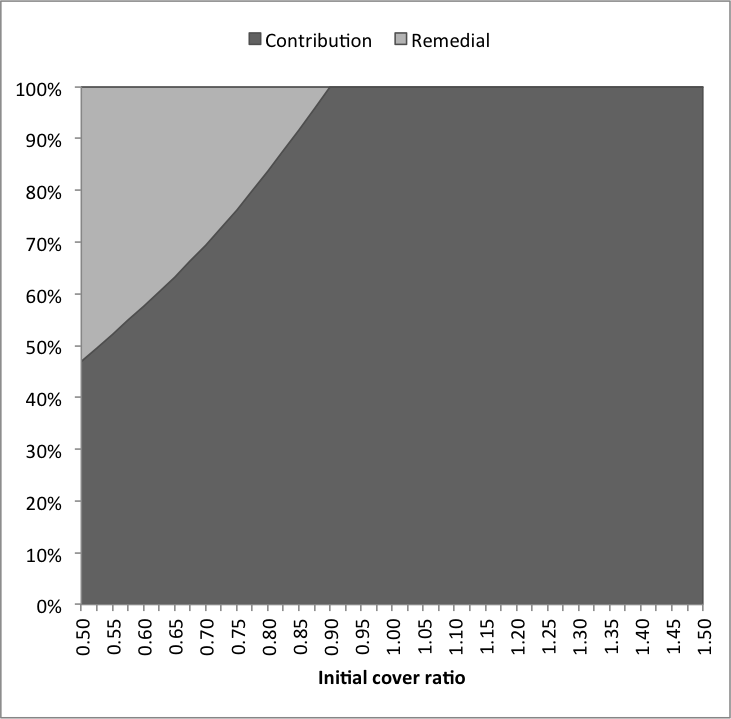}\vspace{0.cm}
		\caption{OICC: Initial cost allocation in function of $F_0$}
		\label{OICC_CR_contrem}
	\end{minipage}
\end{figure}

\begin{figure}[h!]
	\begin{center}
		\includegraphics[width=12cm,height=7cm]{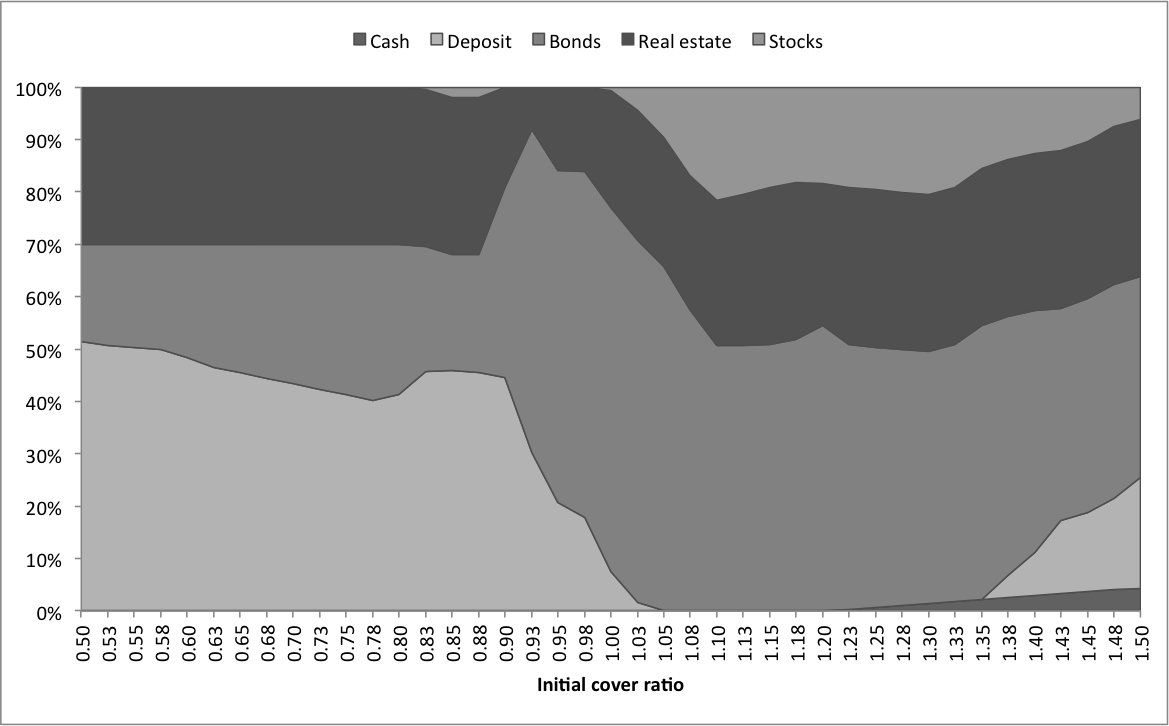} \caption{\label{OICC_CR_allocation}OICC: Asset allocation at $t=0$ in function of $F_0$} 
	\end{center}
\end{figure}
~\\Next we consider the effect of the initial funding ratio on the first stage optimal decision. We vary the value of $L_0$ so that the initial funding ratio $F_0$ lies between $0.5$ and $1.5$, and we assume $\alpha=0.035$. Figure \ref{OICC_CR_crate} displays the evolution of the contribution rate $cr_0$ whereas Figure \ref{OICC_CR_contrem} shows how the total cost is distributed into regular and remedial contributions. As explained earlier, it is conventional to assume that, above a certain funding ratio $F_0^{O^*}$, the remedial contribution is zero and contribution rate decreases as well. According to Figures \ref{OICC_CR_crate} and \ref{OICC_CR_contrem}, the ALM model is set such that $F_0^{O^*}:=0.9$.
\\Figure \ref{OICC_CR_allocation} describes the behaviour of the first stage optimal asset allocation with respect to $F_0$. When $F_0 < F_0^{O^*}$, the optimal asset allocation is stable: approximatively $30\%$ in riskier assets. From $F_0=F_0^{O^*}$, the cover ratio is high enough to no more obtain remedial contribution and to reduce the contribution rate. However, the decision maker has to act in a riskier way in order to meet pension fund liabilities. As a result, the risk exposition increases up to $50\%$ at $F_0=1.275$. An important target of our model is to guarantee a funding ratio $F_5 \geq \bar{F}$ by minimizing the total cost and risk level. For larger values of $F_0$, with a higher chance to fulfill the condition $F_5\geq\bar{F}$, the contribution rate and the risk exposition (i.e. proportion of higher risk assets) decrease. We recall that the objective is not to maximize the wealth, but to minimize the total cost. Thus, the wealthier the fund is, the more prudent the allocation will be.

\subsubsection{MICC}
\begin{figure}[h!] 
	\begin{minipage}[c]{.46\linewidth} 
		\includegraphics[width=7.5cm,height=6.5cm]{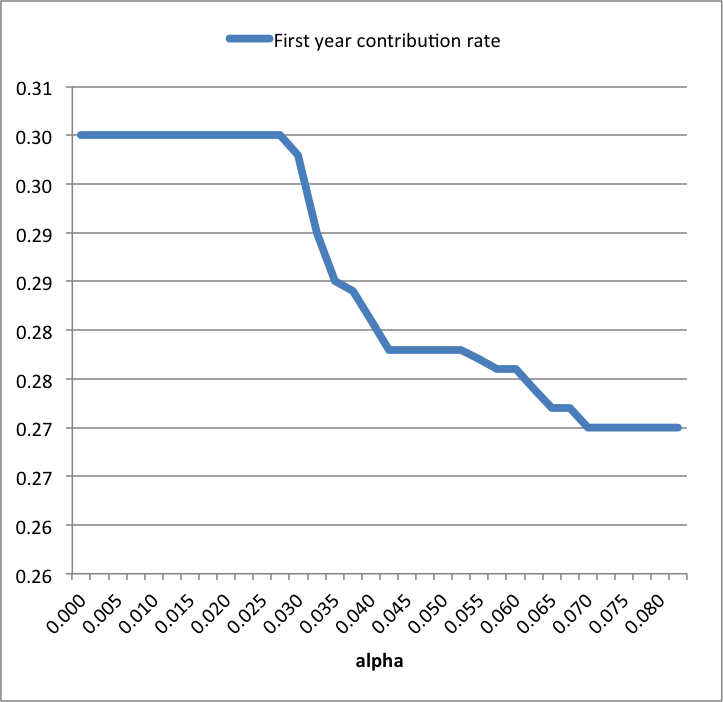}\vspace{0.cm} 
		\caption{MICC: Contribution rate at $t=0$ in function of $\alpha$} 
		\label{MICC_alpha_crate} 
	\end{minipage} \hfill
	\begin{minipage}[c]{.46\linewidth} 
		\includegraphics[width=7.5cm,height=6.5cm]{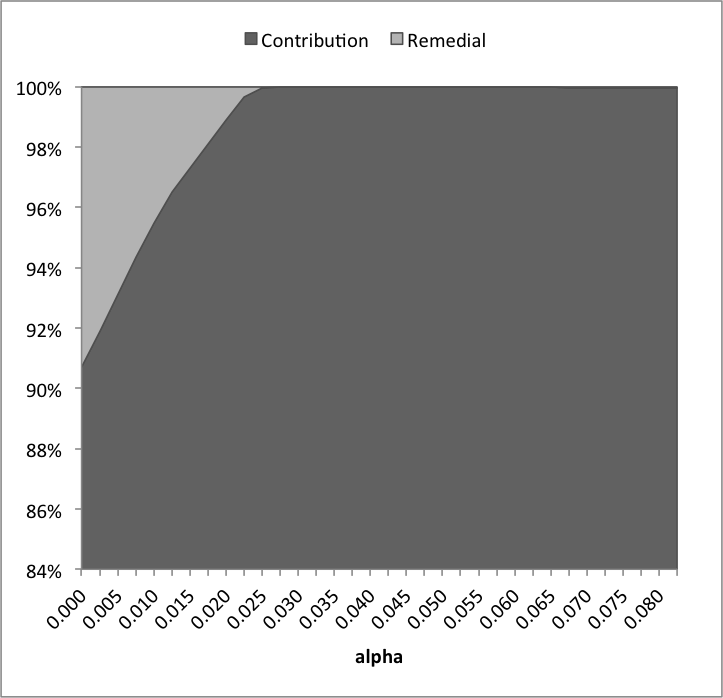}\vspace{0.cm}
		\caption{MICC: Initial cost allocation in function of $\alpha$}
		\label{MICC_alpha_contrem}
	\end{minipage}
\end{figure}

\begin{figure}[h!]
	\begin{center}
		\includegraphics[width=12cm,height=7cm]{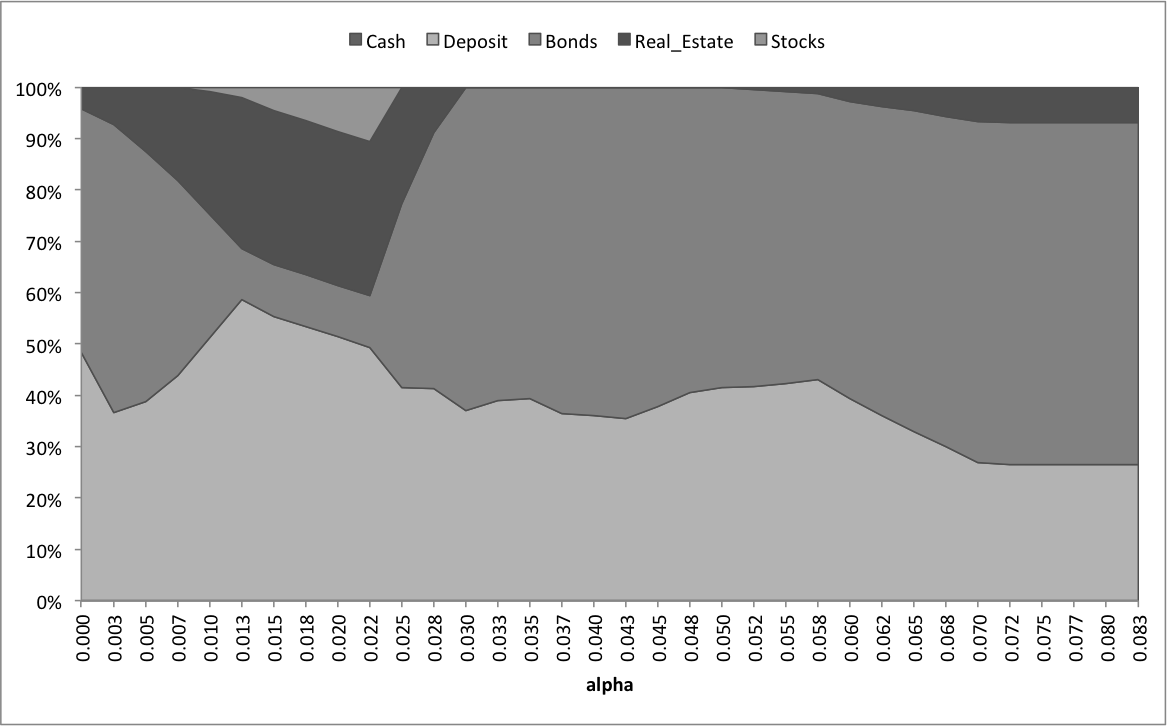} \caption{\label{MICC_alpha_allocation}MICC: Asset allocation at $t=0$ in function of $\alpha$} 
	\end{center}
\end{figure}

\begin{figure}[h!] 
	\begin{minipage}[c]{.46\linewidth} 
		\includegraphics[width=7.5cm,height=6.5cm]{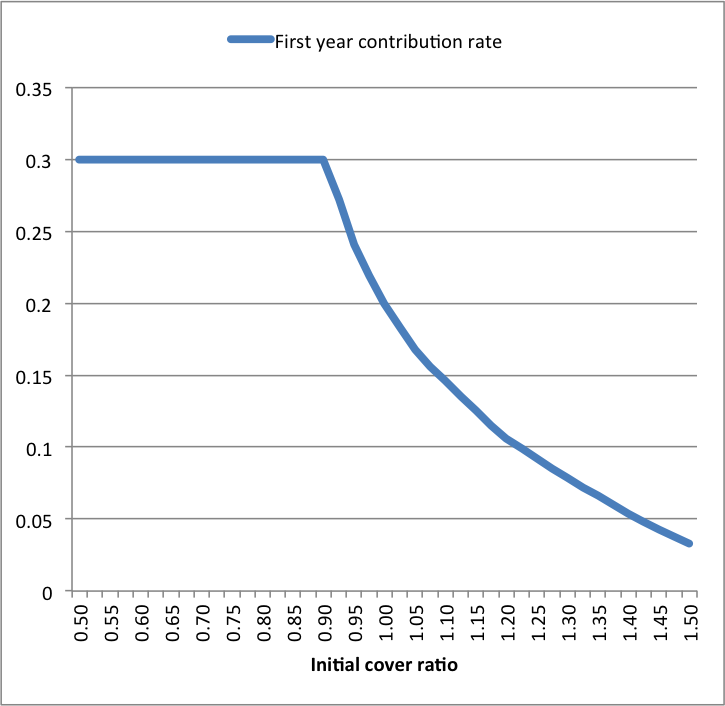}\vspace{0.cm} 
		\caption{MICC: Contribution rate at $t=0$ in function of $F_0$} 
		\label{MICC_CR_crate} 
	\end{minipage} \hfill
	\begin{minipage}[c]{.46\linewidth} 
		\includegraphics[width=7.5cm,height=6.5cm]{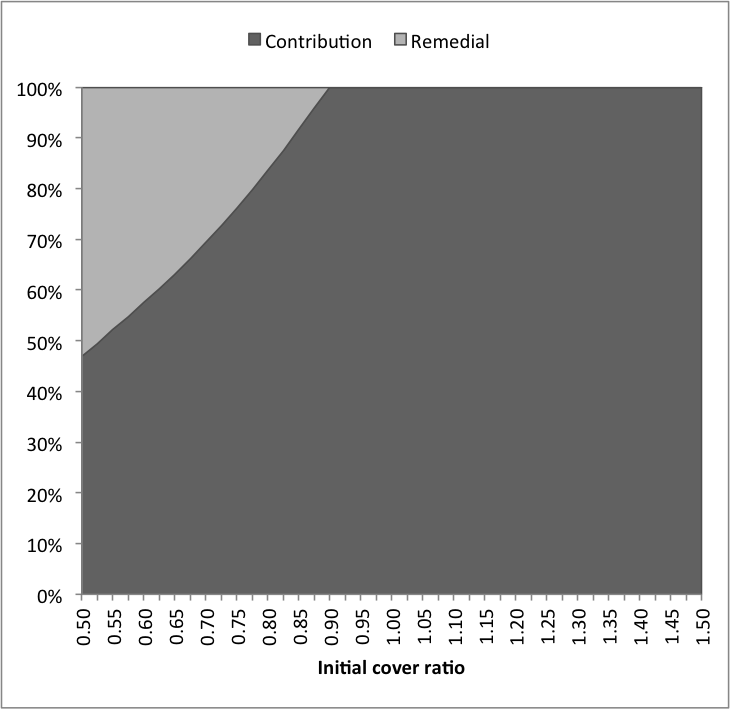}\vspace{0.cm}
		\caption{MICC: Initial cost allocation in function of $F_0$}
		\label{MICC_CR_contrem}
	\end{minipage}
\end{figure}

\begin{figure}[h!]
	\begin{center}
		\includegraphics[width=12cm,height=7cm]{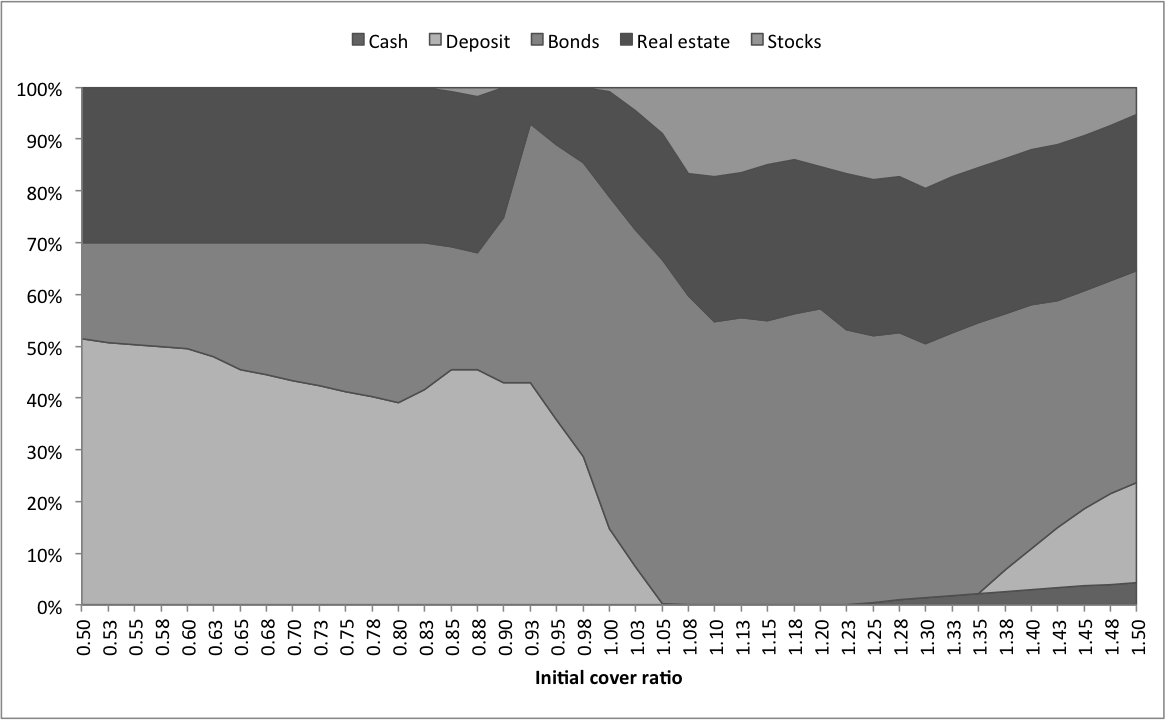} \caption{\label{MICC_CR_allocation}MICC: Asset allocation in function of $F_0$} 
	\end{center}
\end{figure}

In this section, we firstly present the results of the ALM model with the MICC, and secondly, we compare with the OICC. 
For the MICC analysis, assumptions are similar to the ones made for the model with OICC. Figures \ref{MICC_alpha_crate} to \ref{MICC_CR_allocation} display the results of the analysis. The effect of  the risk parameter $\alpha$ is measured in Figures \ref{MICC_alpha_crate}, \ref{MICC_alpha_contrem} and \ref{MICC_alpha_allocation} whereas Figures \ref{MICC_CR_crate}, \ref{MICC_CR_contrem} and \ref{MICC_CR_allocation} analyze the initial funding ratio impact. Although the first stage optimal decisions are different, they behave similarly. The parameters $\alpha_*^M$, $\bar{\alpha}^M$ and $F_0^{M^*}$ (respectively $\alpha_*^O$, $\bar{\alpha}^O$ and $F_0^{O^*}$ for OICC) slightly differ:
\begin{equation}
\alpha_*^M:=0.027;\;\;\;\;\;\;\;\;\;\bar{\alpha}^M:=0.07;\;\;\;\;\;\;\;\;\;F_0^{M^*}:=0.9.
\nonumber
\end{equation}
\begin{figure}[h!] 
	\begin{minipage}[c]{.46\linewidth} 
		\includegraphics[width=7.5cm,height=6.5cm]{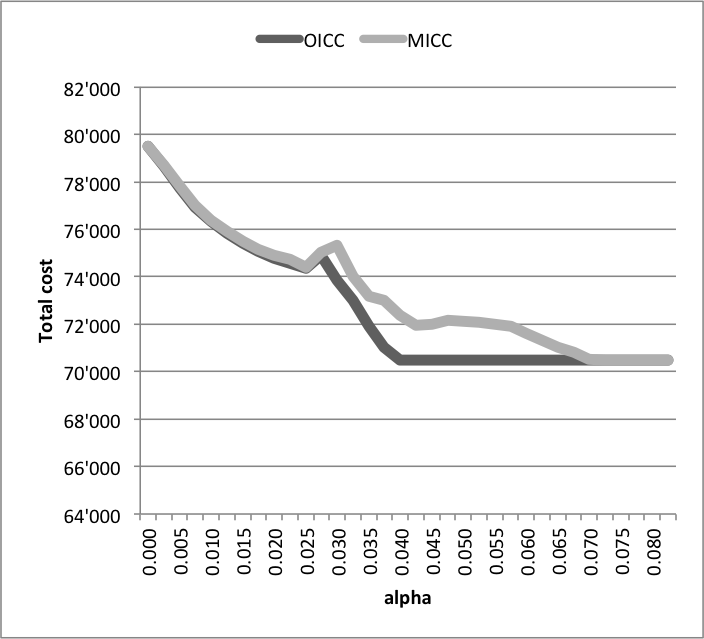}\vspace{0.cm} 
		\caption{Comparison of OICC and MICC in function of $\alpha$: Total cost} 
		\label{MvsO_alpha_cost} 
	\end{minipage} \hfill
	\begin{minipage}[c]{.46\linewidth} 
		\includegraphics[width=7.5cm,height=6.5cm]{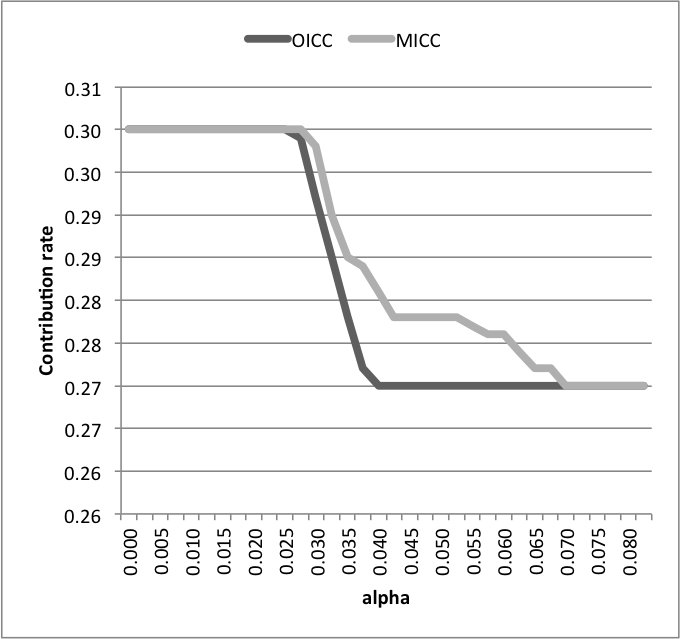}\vspace{0.cm}
		\caption{Comparison of OICC and MICC in function of $\alpha$: Contribution rate $cr_0$}
		\label{MvsO_alpha_crate}
	\end{minipage}
\end{figure}
\begin{figure}[h!] 
	\begin{minipage}[c]{.46\linewidth} 
		\includegraphics[width=7.5cm,height=6.5cm]{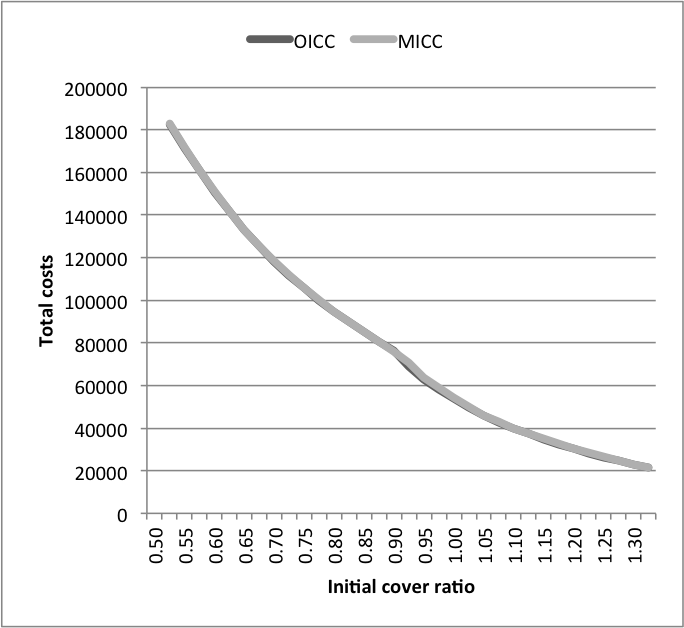}\vspace{0.cm} 
		\caption{Comparison of OICC and MICC in function of $F_0$: Total cost} 
		\label{MvsO_CR_cost} 
	\end{minipage} \hfill
	\begin{minipage}[c]{.46\linewidth} 
		\includegraphics[width=7.5cm,height=6.5cm]{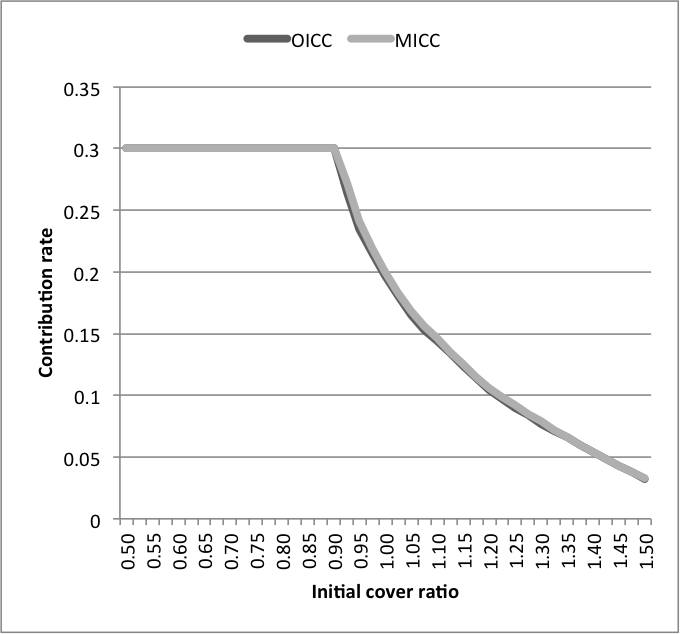}\vspace{0.cm}
		\caption{Comparison of OICC and MICC in function of $F_0$: Contribution rate $cr_0$}
		\label{MvsO_CR_crate}
	\end{minipage}
\end{figure}
~\\The MICC approach is basically more rigid than OICC. That is why the above parameters are greater or equal to their analogues in OICC. Notice that, according to Figures \ref{OICC_alpha_allocation} and \ref{MICC_alpha_allocation} and for $\alpha\geq \bar{\alpha}^O$ in OICC and $\alpha\geq \bar{\alpha}^M$ in MICC, the asset allocations are \textit{exactly} the same; showing that when $\alpha$ is above $\bar{\alpha}^O$ (resp. $\bar{\alpha}^M$), the OICC (resp. MICC) has no influence on the initial ALM model. In general, the optimal decisions obtained from OICC and MICC slightly differ. For example, when $\alpha=0.05$ and $F_0=0.9166$, the first stage optimal decision of the MICC is:
\begin{equation}
H_0:= \begin{pmatrix}
0.41 & 0.59 & 0 & 0
\end{pmatrix}^\top ; \;\;\;\;\;\;\;\;\; cr_0=0.279\;\;\;\;\;\;\;\;\; \text{and} \;\;\;\;\;\;\;\;\;Z_0=0
\nonumber
\end{equation}
whereas for the OICC:
\begin{equation}
H_0:= \begin{pmatrix}
0.26 & 0.66 & 0.08 & 0
\end{pmatrix}^\top ; \;\;\;\;\;\;\;\;\; cr_0=0.270;\;\;\;\;\;\;\;\; \text{and} \;\;\;\;\;\;\;\;\;Z_0=0.
\nonumber
\end{equation}
Asset allocations are calculated as percentage of the total asset. According to the above example, the decisions related to the OICC approach are riskier than the ones of MICC, especially regarding the asset allocation. In what follows, we will try to quantify the cost of this risk reduction. For a pension fund, this can be done by measuring the difference in term of the total cost (regular contribution $+$ remedial contribution). Hence, Figure  \ref{MvsO_alpha_cost} compares the total costs of OICC and MICC whereas Figure \ref{MvsO_alpha_crate} displays the contribution rate difference, all this with respect to $\alpha$. When $\alpha\leq \alpha_*^O=0.025$ or $\alpha\geq \bar{\alpha}^M=0.07$, the contribution rate and total cost are equal for both models. For $\alpha_*^O\leq\alpha\leq \bar{\alpha}^M$, OICC and MICC slightly differ, i.e. MICC costs more for a maximum variation of $2'000$ (less than $2\%$ of total asset) and $1.5\%$, respectively, for the total cost and the contribution rate. Consequently, although being more conservative, the MICC is preferable the OICC for the two following reasons:
\begin{itemize}
\item it is safer, and 
\item the cost of this safety is not significant: less than $2\%$ of total asset.
\end{itemize}
Moreover, Figures \ref{MvsO_CR_cost} and \ref{MvsO_CR_crate}, which compare the effect of $F_0$ on the OICC and on the MICC, confirm that result.

\section{Conclusion} \label{conclusion}
In this paper, we considered the effects of integrated chance constraints (ICC) on an ALM model for a defined benefit pension fund. ICC are appropriate for modeling risk constraints, in particular when a quantitative risk measure is preferable. At each decision time, they put an upper bound on the one period expected shortfalls. The upper limit considered is a fixed proportion of total liability making the risk parameter scale free as well as time dependent. ICC are very attractive from a computational perspective. Their impact on the ALM model is analyzed through the implementation of a multistage stochastic linear program. We defined two types of ICC: the one period integrated chance constraint (OICC) and the multistage integrated chance constraint (MICC).
\\The first step of the work consisted on describing the influences of the ICCs on the time $0$ optimal decisions. This impact is measured by the way of the risk parameter and of the initial cover ratio. As one could expect, the risk exposition of the optimal portfolio tends to increase with the risk parameter until this latter reaches a certain value, from which, the allocation remains stable thereafter. On the other hand, for reasonable values of the initial cover ratio, the risk exposition is increasing. Above a certain value, as there is a lower chance of not meeting the target funding ratio condition and in order to guarantee the benefit payments at a low cost, the risk exposition decreases.
\\MICC is basically more restrictive and more cautious than OICC. Secondly, the cost generated by the rise in security with the MICC is quantified. Although the optimal decisions from the OICC and the MICC are not the same, the total costs are very close, showing that the MICC is definitely a better approach.
\\However, the obtained results are subject to discussion since they mainly rely on the scenario tree. In further considerations, it would be interesting to analyse the result stability regarding the scenario tree.\\
\section*{Acknowledgment}
The authors would like to thank Stefan Thonhauser (TU-Graz, Austria) and Joel Wagner (Department of Actuarial Science, HEC, University of Lausanne) for their time, their advice, their technical supports and all the interesting discussions we had during the preparation of this contribution.


\bibliographystyle{plain}

\section*{APPENDICES}
\subsection*{Appendix 1: The ALM program description}

We start this section by defining indices, variables and parameters of the model. Secondly, the ALM model with the objective and the constraints are also displayed.\\
\\\textit{\textbf{Indices}}
\begin{flalign*}
 t & \;\;\;\;\;\;\;\; \text{time index},\; t=0,1,\cdots ,T\\
 s & \;\;\;\;\;\;\;\; \text{scenarios index},\; s=1,\cdots ,S\\
 k & \;\;\;\;\;\;\;\; \text{index of asset classes},\; k=1,\cdots ,d
\end{flalign*}
\textit{\textbf{Decision variables}}
\begin{flalign*}
 Z_t^s  & \;\;\;\;\;\;\;\; \text{remedial contribution by the sponsor at time } t \text{ in scenario } s \\
 C_t^s  & \;\;\;\;\;\;\;\; \text{total cash amount at the beginning of year } t \text{ in scenario } s\\
 H_{k,t}^s & \;\;\;\;\;\;\;\; \text{value of the investments hold in asset class } k, \text{ at the beginning of year } t \text{ and in scenario } s\\
 B_{k,t}^s & \;\;\;\;\;\;\;\; \text{value of the asset class } k, \text{ bought at the beginning of year } t \text{ and in scenario } s\\
 S_{k,t}^s & \;\;\;\;\;\;\;\; \text{value of the asset class } k, \text{ sold at the beginning of year } t \text{ and in scenario } s\\
 cr_t^s & \;\;\;\;\;\;\;\; \text{contribution rate for year } t+1 \text{ in scenario } s\\
 A_t^s & \;\;\;\;\;\;\;\; \text{total asset value at time } t \text{ in scenario } s\\
 A_t^{*s} & \;\;\;\;\;\;\;\; \text{total asset value just before the asset allocation and the remedial contribution at time } t \text{ in scenario } s\\
 \Delta_{cr_t}^s & \;\;\;\;\;\;\;\; \text{variation (increase or decrease) of contribution rate from year } t \text{ to } t+1 \text{ in scenario } s
\end{flalign*}
\textit{\textbf{Random parameters}}
\begin{flalign*}
 r_{k,t}^s & \;\;\;\;\;\;\;\; \text{random rate of return on asset class } k \text{ over year } t \text{ in scenario } s \text{ and } \xi_{k,t}:=1+r_{k,t}\\
 W_t^s & \;\;\;\;\;\;\;\; \text{random total wages of active participants in year } t \text{ in scenario } s\\
 L_t^s & \;\;\;\;\;\;\;\; \text{random value of liabilities at time } t \text{ in scenario } s\\
 {Ben}_t^s & \;\;\;\;\;\;\;\; \text{random total benefit payments to active participants in year } t \text{ in scenario } s
\end{flalign*}
\textit{\textbf{Deterministic parameters}}
\begin{flalign*}
 T & \;\;\;\;\;\;\;\; \text{time horizon}\\
 S & \;\;\;\;\;\;\;\; \text{number of scenarios}\\
 d & \;\;\;\;\;\;\;\; \text{number of asset classes}\\
 \alpha & \;\;\;\;\;\;\;\; \text{risk parameter defined by either the sponsor or the regulator}\\
 \bar{c}_k^B & \;\;\;\;\;\;\;\; \text{proportional transaction cost for purchasing an asset class } k\\
 \bar{c}_k^S & \;\;\;\;\;\;\;\; \text{proportional transaction cost for purchasing an asset class } k\\
 l_k & \;\;\;\;\;\;\;\; \text{lower bound on the proportion of asset class } k \text{in the total asset portfolio}\\
 u_k & \;\;\;\;\;\;\;\; \text{upper bound on the proportion of asset class } k \text{in the total asset portfolio}\\
 l_c & \;\;\;\;\;\;\;\; \text{lower bound on the proportion of cash } k \text{in the total asset portfolio}\\
 u_c & \;\;\;\;\;\;\;\; \text{upper bound on the proportion of cash } k \text{in the total asset portfolio}\\
 {cr}^l & \;\;\;\;\;\;\;\; \text{lower bound on the contribution rate}\\
 {cr}^u & \;\;\;\;\;\;\;\; \text{upper bound on the contribution rate}\\ 
 \underline{\Delta_{cr}} & \;\;\;\;\;\;\;\; \text{lower bound on the yearly absolute variation of the contribution rate}\\
 \bar{\Delta}_{cr} & \;\;\;\;\;\;\;\; \text{upper bound on the yearly absolute variation of the contribution rate}\\
 \bar{F} & \;\;\;\;\;\;\;\; \text{target funding ratio}\\
 \gamma & \;\;\;\;\;\;\;\; \text{lower bound on the funding ratio}\\
 r_f & \;\;\;\;\;\;\;\; \text{risk free interest rate}\\
 \lambda_z & \;\;\;\;\;\;\;\; \text{penalty parameter for remedial contribution}\\
 \lambda_{\Delta_{cr}} & \;\;\;\;\;\;\;\; \text{penalty parameter for absolute variation of contribution rate}\\
 \bar{H}_{k,0} & \;\;\;\;\;\;\;\; \text{value of the initial allocation in asset class } k\\
 \bar{C}_0  & \;\;\;\;\;\;\;\; \text{initial cash amount}
\end{flalign*}
\textit{\textbf{Objective}}
\\The objective of the model is to determine the asset allocation, contribution rate and remedial contributions that minimize the total cost defined as follows
\begin{equation*}
\min_{H,cr,Z} \mathbb{E}_{0}\left[  \sum_{t=0}^{T-1} v_{t+1} \left( cr_tW_{t+1} + \lambda_z Z_{t+1}  \right) + \sum_{t=0}^{T-2} v_{t+1}\lambda_{\Delta_{cr}}  {\Delta_{{cr}_t}}  W_{t+1} \right]
\end{equation*}
under the following constraints. In the model description, anytime we use indices $s$ and/or $k$ is equivalent to saying for any $s=1,\cdots,S$ and/or for any $k=1,\cdots,d$.\\
\\\underline{\textit{Budget constraints and total value of the assets}}
\begin{flalign*}
 A_{0} & =\sum_{k=1}^{d}\bar{H}_{k,0} + \bar{C}_{0} + Z_0  - \sum_{k=1}^{d} \left( \bar{c}_k^BB_{k,0}+\bar{c}_k^SS_{k,0}\right)=\sum_{k=1}^{d}H_{k,0} + C_{0}\\
 A_{T}^s & =\sum_{k=1}^{d}H_{T-1,k}^s \xi_{T,k}^s + C_{T-1}^s \left( 1+r_f\right)  + cr_{T-1}^s W_T^s-\text{Ben}_T^s=A_T^{*s}\\
 A_{t}^s & =\sum_{k=1}^{d}H_{k,t-1}^s\xi_{k,t}^s + C_{t-1}^s\left( 1+r_f\right)  + cr_{t-1}^s W_t^s - \text{Ben}_t^s + Z_t^s  - \sum_{k=1}^{d} \left( \bar{c}_k^BB_{k,t}^s+\bar{c}_k^SS_{k,t}^s\right); \;\;\;\;\; t =  1,\cdots,T-1\\
 A_t^s & =A_t^{*s} + Z_t^s  - \sum_{k=1}^{d} \left( \bar{c}_k^BB_{k,t}^s+\bar{c}_k^SS_{k,t}^s\right)=\sum_{k=1}^{d}H_{k,t}^s + C_{t}^s \; ; \;\;\;\;\; t =  1,\cdots,T-1
\end{flalign*}
\\\underline{\textit{Asset classes dynamics}}
\begin{flalign*}
H_{k,0} & =\bar{H}_{k,0}+B_{k,0}-S_{k,0}\\
H_{k,t}^s & =\xi_{k,t}^sH_{k,t-1}^s+B_{k,t}^s-S_{k,t}^s \;; \;\;\;\;\; t =  1,\cdots,T 
\end{flalign*}
\\\underline{\textit{Cash dynamics}}
\begin{flalign*}
C_0 & =\bar{C}_0 + Z_0 -  \sum_{k=1}^{d} \left(1+\bar{c}^B_k \right)B_{k,0} + \sum_{k=1}^{d} \left(1-\bar{c}^S_k \right) S_{k,0}\\
C_t^s & =C_{t-1}^s\left(1 +r_f \right) + cr_{t-1}^s W_t^s - \text{Ben}_t^s + Z_t^s - \sum_{k=1}^{d} \left(1+\bar{c}^B_k \right)B_{k,t}^s + \sum_{k=1}^{d} \left(1-\bar{c}^S_k \right) S_{k,t}^s \;; \;\;\;\;\; t =  1,\cdots,T
\end{flalign*}
\\\underline{\textit{Not short selling assets and not borrowing cash constraints}}
\begin{flalign*}
H_{k,t}^s\geq0\;; \;\;B_{k,t}^s\geq 0\;; \;\;S_{k,t}^s\geq 0\;; \;\;C_{t}^s\geq 0\;; \;\;\;\;\; t =  0,\cdots,T
\end{flalign*}
\\\underline{\textit{Liquidity constraints}}
\begin{flalign*}
C_t^s \left( 1+r_f \right) + \mathbb{E}_{t,s}\left( cr_t^s W_{t+1} - Ben_{t+1} \right) \geq0\;; \;\;\;\;\; t =  0,\cdots,T-1
\end{flalign*}
\\\underline{\textit{Portfolio constraints}}
\begin{flalign*}
& l_kA_t^s  \leq H_{k,t}^s\leq u_kA_t^s \; ; \;\;\;\;\; t =  0,\cdots,T-1\\
& l_c A_t^s \leq C_t^s \leq u_c A_t^s \; ; \;\;\;\;\; t =  0,\cdots,T-1
\end{flalign*}
\\\underline{\textit{Constraints on contribution rates}}
\begin{flalign*}
{cr}^l \leq {cr}_t^s \leq {cr}^u \;\; \text{  and  }  \;\; \underline{\Delta_{cr}} \leq \Delta_{{cr}_t}^s \leq \bar{\Delta}_{cr} \; ; \;\;\;\;\; t =  0,\cdots,T-1
\end{flalign*}
The decision variables are subject to the non-anticipativity constraints. The integrated chance constraints defined in section \ref{RiskConstraint} control the risk-level of the model. They also have to be included in the model as explained in section \ref{ICC_interpretation}.

\subsection*{Appendix 2: Proposition \ref{my_proposition}, an example based sketch of proof}
We consider the event tree of Figure \ref{arbre_scenario_appendix} with a time horizon $T=3$ and a branching structure of $1-5-4-2$, leading to $S=5\times4\times2=40$ scenarios. A node is a possible outcome of the stochastic event at a given time. The starting and ending nodes of the tree are round whereas the others are rectangular. At each $t\in\left\lbrace 2,\;3\right\rbrace $, the rectangular nodes are name according to time and following a top-down alphabetic order. For example, the rectangle $\left( 1,a \right) $ describe the outcome at the first node of time $1$. At each node, the economical values such as total asset, total liability and expected shortfall can be determined. For sake of clarity, we recall that, due to non-anticipativity, the node $\left( 1,a \right) $ is equivalent to the node $\left( 1,s' \right), \;s'\in\left\lbrace 1,\cdots,8\right\rbrace $ as described before, the node $\left( 1,b \right) $ is equivalent to the node $\left( 1,s' \right), \;s'\in\left\lbrace 9,\cdots,16\right\rbrace $ and so on for the other nodes. The other notations used here are similar to the ones in the paper.
\begin{figure}[t!]
\centering
\begin{tikzpicture}[grow=right,->,>=stealth'] 
\node [arn_n] {0}
    child{ node [arn_x] {\tiny 1,e} 
            child{ node [arn_x] {\tiny 2,t}
							child{ node [arn_n] {{\tiny 40}}}
							child{ node [arn_n] {{\tiny 39}}}
            } 
            child{ node [arn_x] {\tiny 2,s}}
            child{ node [arn_x] {\tiny 2,r}}
            child{ node [arn_x] {\tiny 2,q} 
            	        child{ node [arn_n] {{\tiny 34}} edge from parent node[above left]
                         {}} 
							child{ node [arn_n] {{\tiny 33}}}
            }                                       
    }
    child{ node [arn_x] {\tiny 1,d}}
    child{ node [arn_x] {\tiny 1,c}}
    child{ node [arn_x] {\tiny 1,b}}
    child{ node [arn_x] {\tiny 1,a}
            child{ node [arn_x] {\tiny 2,d}
                    child{ node [arn_n] {{\tiny 8}}}
				    child{ node [arn_n] {{\tiny 7}}}} 
            child{ node [arn_x] {\tiny 2,c}}
            child{ node [arn_x] {\tiny 2,b}}
            child{ node [arn_x] {\tiny 2,a}
							child{ node [arn_n] {{\tiny 2}}}
							child{ node [arn_n] {{\tiny 1}}}
            }
		}
; 
\end{tikzpicture}\\
${\tiny t=0}\;\;\;\;\;\;\; \text{event} \;\;\;\;\;\;\;{\tiny t=1}\;\;\;\;\; \text{event} \;\;\;\;\;{\tiny t=2}\;\;\;\;\; \text{event} \;\;\;\;\;{\tiny t=3}$
\caption{\label{arbre_scenario_appendix}A scenario tree with $40$ scenarios and $66$ nodes.}
\end{figure}
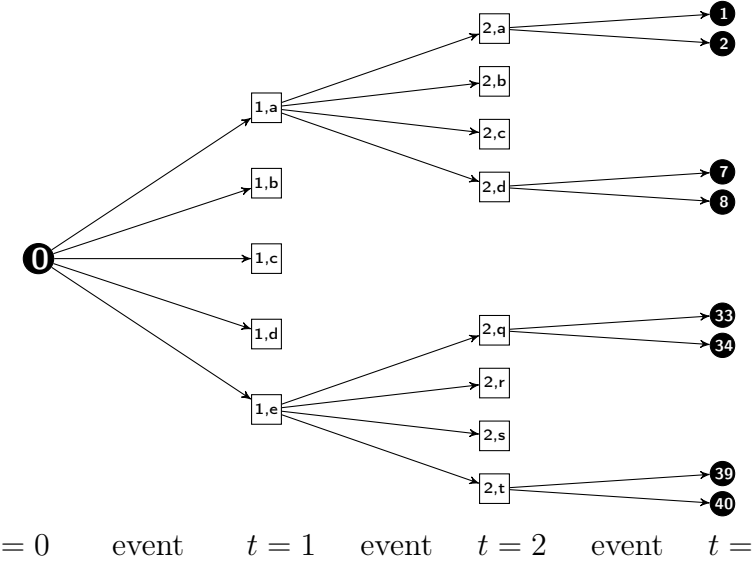We define
\begin{equation}
\Lambda_{\left( t,s \right) }:=\sum_{s^{'} \in \mathcal{S}} p_{t,s}^{s^{'}} \left( A_{t+1}^{*s^{'}} - \gamma L_{t+1}^{s^{'}} \right)^{-}
\nonumber
\end{equation}
where $p_{t,s}^{s^{'}}$ stands for the conditional probability to reach node $\left(t+1,s^{'}\right)$ going from $\left( t,s\right)$ and $p_{t,s}^{s^{'}}=0$ for any scenario $s^{'}$ of $t+1$ not descending from $\left( t,s\right) $. The MICC defined in equation \ref{chance_prime1} is then
\begin{equation}
\left\lbrace \Lambda_{h-1,s}, \;\;h\in \mathcal{T}_{t+1}  \right\rbrace  \leq \beta_{t,s}, \;\; t\in \left\lbrace 0,\;1,\;2 \right\rbrace.
\nonumber
\end{equation}
The value $\beta_{t,s}:=\alpha L_t^s$ is the ICC upper limit computed at time $t$ in scenario $s$. According to this set of constraints,\\
\\\underline{from initial node at $t:=0$}
\begin{equation}
h \in \left\lbrace 1,\;2,\;3\right\rbrace \Rightarrow \Lambda_{0,s} \leq \beta_0, \; \Lambda_{1,s} \leq \beta_0, \; \text{and} \; \Lambda_{2,s} \leq \beta_0, \;\;\; s\in \mathcal{S}.
\end{equation}
\underline{At $t:=1$},
\begin{align*}
\text{from node }(1,a), \; & h \in \left\lbrace 2,\;3\right\rbrace \Rightarrow \Lambda_{1,s} \leq \beta_{1,s}, \; \text{and} \; \Lambda_{2,s} \leq \beta_{1,s}, \;\;\; s\in \left\lbrace 1,\; 2, \cdots, \;8\right\rbrace \\
\text{from node }(1,b), \; & h \in \left\lbrace 2,\;3\right\rbrace \Rightarrow \Lambda_{1,s} \leq \beta_{1,s}, \; \text{and} \; \Lambda_{2,s} \leq \beta_{1,s}, \;\;\; s\in \left\lbrace 9,\; 10, \cdots, \;16\right\rbrace \\
\vdots \;\;\;\;\;\;\;\;\;\; & \;\;\;\;\;\;\;\;\;\; \;\;\;\;\;\;\;\;\;\;\vdots\\
\vdots \;\;\;\;\;\;\;\;\;\; & \;\;\;\;\;\;\;\;\;\; \;\;\;\;\;\;\;\;\;\;\vdots\\
\text{from node }(1,e), \; & h \in \left\lbrace 2,\;3\right\rbrace \Rightarrow \Lambda_{1,s} \leq \beta_{1,s}, \; \text{and} \; \Lambda_{2,s} \leq \beta_{1,s}, \;\;\; s\in \left\lbrace 33,\; 34, \cdots, \;40\right\rbrace.
\end{align*}
\underline{At $t:=2$}
\begin{align*}
\text{from node }(2,a), \; & h=3 \Rightarrow \; \text{and} \; \Lambda_{2,s} \leq \beta_{2,s}, \;\;\; s\in \left\lbrace 1,\; 2\right\rbrace \\
\text{from node }(2,b), \; & h=3 \Rightarrow \; \text{and} \; \Lambda_{2,s} \leq \beta_{2,s}, \;\;\; s\in \left\lbrace 3,\; 4\right\rbrace \\
\vdots \;\;\;\;\;\;\;\;\;\; & \;\;\;\;\;\;\;\;\;\; \;\;\;\;\;\;\;\;\;\;\vdots\\
\vdots \;\;\;\;\;\;\;\;\;\; & \;\;\;\;\;\;\;\;\;\; \;\;\;\;\;\;\;\;\;\;\vdots\\
\text{from node }(2,t), \; & h=3 \Rightarrow \; \text{and} \; \Lambda_{2,s} \leq \beta_{2,s}, \;\;\; s\in \left\lbrace 39,\; 40\right\rbrace.
\end{align*}
Therefore, we obtain for any $s\in \mathcal{S}$ that
\begin{align*}
\text{at time }0, \; & \Lambda_{0,s} \leq \beta_0  &  \Leftrightarrow \Lambda_{0,s} \leq \beta_0\\
\text{at time }1, \; & \Lambda_{1,s} \leq \beta_0,\; \Lambda_{1,s} \leq \beta_{1,s}   &  \Leftrightarrow \Lambda_{1,s} \leq \min\left\lbrace \beta_0, \;\beta_{1,s} \right\rbrace \\
\text{at time }2, \; & \Lambda_{2,s} \leq \beta_0,\; \Lambda_{2,s} \leq \beta_{1,s}, \; \Lambda_{2,s} \leq \beta_{1,s}   &  \Leftrightarrow \Lambda_{2,s} \leq \min\left\lbrace \beta_0, \;\beta_{1,s}, \;\beta_{2,s}  \right\rbrace.
\end{align*}
The obtained result leads obviously to the proposition \ref{my_proposition} in the paper. Considering such an example is therefore without loss of generality. For an other event tree with different time horizon and branching structure, the proposition can be prooved using the same procedure.

\end{document}